\documentclass[sigconf]{acmart}

\usepackage{graphicx}
\usepackage{caption}
\usepackage{subcaption}
\usepackage{multirow}
\usepackage{multicol}
\usepackage{colortbl}

\AtBeginDocument{%
  \providecommand\BibTeX{{%
    \normalfont B\kern-0.5em{\scshape i\kern-0.25em b}\kern-0.8em\TeX}}}

\setcopyright{acmcopyright}

\copyrightyear{2020}
\acmYear{2020}
\setcopyright{none}
\acmConference[MSR '20]{17th International Conference on Mining Software Repositories}{October 5--6, 2020}{Seoul, Republic of Korea}
\acmBooktitle{17th International Conference on Mining Software Repositories (MSR '20), October 5--6, 2020, Seoul, Republic of Korea}
\acmPrice{15.00}
\acmDOI{10.1145/3379597.3387446}
\acmISBN{978-1-4503-7517-7/20/05}

\hypersetup{draft}
\begin{document}

\title{Can We Use SE-specific Sentiment Analysis Tools in a Cross-Platform Setting?}

\author{Nicole Novielli, Fabio Calefato, Davide Dongiovanni, Daniela Girardi, Filippo Lanubile}
\affiliation{University of Bari, Italy}
\email{nicole.novielli,fabio.calefato,daniela.girardi,filippo.lanubile@uniba.it}
\email{d.dongiovanni@studenti.uniba.it}








\renewcommand{\shortauthors}{Novielli, et al.}

\begin{abstract}
  In this paper, we address the problem of using sentiment analysis tools ‘off-the-shelf’, that is when a gold standard is not available for retraining. We evaluate the performance of four SE-specific tools in a cross-platform setting, i.e., on a test set collected from data sources different from the one used for training. We find that (i) the lexicon-based tools outperform the supervised approaches retrained in a cross-platform setting and (ii) retraining can be beneficial in within-platform settings in the presence of robust gold standard datasets, even using a minimal training set. Based on our empirical findings, we derive guidelines for reliable use of sentiment analysis tools in software engineering.
\end{abstract}


\maketitle

\section{Introduction}
\label{sec:introduction}
Investigating the role of affect has emerged as a consolidated trend of research on human aspects in software engineering ~\cite{IEEESoftwareEmotions2019,NOVIELLI2019JSS}. Sentiment analysis is used to detect emotions in social coding platforms, such as GitHub~\cite{Guzman14,Pletea14,Sinha16}, issue-tracking tools, such as Jira~\cite{Gachechiladze17,Mantyla16,Ortu15}, and collaborative knowledge-sharing sites, such as Stack Overflow~\cite{CalefatoIST18}. Further research has been leveraging sentiment analysis for requirements elicitation based on opinion detection in user-generated content~\cite{Guzman16,Maalej17}. 

Despite the popularity of general-purpose sentiment analysis tools, a consensus has been reached in the research community about the negative results obtained when using such tools ‘off-the-shelf’ to detect developers’ emotions~\cite{Jongeling17,Lin18,Novielli15}, thus indicating the need for fine-tuning such tools for the software engineering domain~\cite{Menzies2020}. Trying to overcome these limitations, researchers have started to build their own classifiers specifically customized for software engineering (SE)~\cite{Blaz16,Chen:fse2019,Gachechiladze17,Islam17,Islam18}. Among others, four SE-specific tools are publicly available, namely Senti4SD~\cite{CalefatoEMSE18}, SentiStrength-SE~\cite{Islam17}, SentiCR~\cite{Ahmed17}, and DEVA~\cite{Islam18}. SentiStrength-SE and DEVA implement a lexicon-based approach and have been optimized on a gold standard dataset of manually annotated content from Jira. Conversely, Senti4SD and SentiCR implement a supervised approach for training polarity classification models. They also offer retraining functions for model optimization and fine-tuning based on a custom gold standard dataset. 

In a previous benchmarking study~\cite{Novielli18}, we compared the predictions of Senti4SD, SentiCR, and SentiStrength-SE, showing how domain-specific customization provides a boost in accuracy compared to the baseline approach represented by SentiStrength~\cite{Thelwall2010}, an off-the-shelf tool trained and validated on general-purpose social media. Specifically, the best performance was observed for tools implementing supervised approaches. Based on this evidence, customized retraining of the classifiers was recommended. We executed the study in a within-platform setting, that is, we trained and tested each classifier using a gold standard from the same data source. However, building a manually annotated gold standard is a time-consuming task and, as such, not always feasible.

In this paper, we address the problem of using SE-specific sentiment analysis tools in a cross-platform setting, i.e., in the absence of a gold standard for a target data source. Our study builds upon the design and results of two previous studies, one by \citeauthor{Jongeling17} \cite{Jongeling17} and one of our previous works \cite{Novielli18}, assessing the performance of general-purpose and SE-specific sentiment analysis tools, respectively. Specifically, in line with these previous studies~\cite{Jongeling17,Novielli18}, we define the following research questions:
\begin{itemize}
    \item $RQ1$ - To what extent do different SE-specific sentiment analysis tools agree with the emotions of software developers when used as ‘off-the-shelf’ tools in a cross-platform setting? 
    \item $RQ2$ - To what extent do results from different SE-specific sentiment analysis tools agree with each other when used as ‘off-the-shelf’ tools in a cross-platform setting?
\end{itemize}

To enable the comparison with previous research, we assess the tool performance on two gold standard datasets from the software engineering domain, namely a Jira dataset of 6K comments ~\cite{Ortu16} and a Stack Overflow dataset of 4K posts (questions, answers, and comments). Both datasets have been manually annotated by adopting a model-driven approach, that is by referring to theoretical emotion models translated into detailed guidelines for the human raters. Specifically, the annotation of the Jira and Stack Overflow dataset is based on the theoretical model of emotions defined by Shaver et al.~\cite{Shaver87}. Furthermore, we create and include in our benchmark a gold standard dataset of 7K comments from GitHub pull-request and commit comments. The GitHub comments have been manually annotated by three of the authors, following the same annotation guidelines adopted for the Stack Overflow and Jira datasets. 

Finally, we aim to understand how many training documents are required so that choosing to retrain a supervised tool is convenient compared to using a lexicon-based approach `off-the-shelf'. Indeed, building a manually annotated gold standard for sentiment analysis is a time-consuming task that requires careful training of the raters and the appropriate choice of the emotion model~\cite{Novielli15}. As such, we formulate a third research question:

\begin{itemize}
    \item $RQ3$ - To what extent is the performance of SE-specific sentiment analysis tools affected by the size of the training set?
\end{itemize}

The contributions of this paper are as follows. First, we release a dataset of more than 7k comments from GitHub.\footnote{The dataset can be downloaded from: \url{https://doi.org/10.6084/m9.figshare.11604597}} 
To the best of our knowledge, this is the first publicly available dataset including texts from GitHub annotated for sentiment polarity. As a second contribution, we enhance the current understanding of the most frequent causes of misclassification due to cross-platform use of sentiment analysis tools when applied in the software engineering domain. Finally,  we derive empirically-based recommendations for the safe adoption of SE-specific tools for sentiment analysis, both in presence and in absence of a gold standard for retraining.   

The paper is organized as follows. In Section~\ref{sec:background}, we address sentiment analysis in software engineering and summarize the previous benchmarking studies we build upon. In Section~\ref{sec:tools}, we describe the SE-specific tools we include in our benchmark. In Section~\ref{sec:datasets}, we describe the three manually annotated gold standard dataset that we include in our benchmark, with a detailed description of the annotation study we conduct to build the GitHub gold standard. Then, we describe the study design in Section~\ref{sec:experimental-setting}, report results in Section~\ref{sec:results}, and present guidelines for sentiment analysis in SE in Section~\ref{sec:discussion}. Finally, we discuss the threats to validity in Section~\ref{sec:threats} and conclude in Section~\ref{sec:conclusions}.

\section{Sentiment Analysis in Software Engineering}
\label{sec:background}
Sentiment analysis is the task of mining the positive or negative opinions and emotions conveyed by text~\cite{Pang07}.
Psychologists have worked for decades at the definition of theoretical model for emotions~\cite{Lazarus1991,russell1980,Shaver87}. Regardless of the specific taxonomy, emotions can be mapped to the polarity dimension, i.e., classified as positive, negative, or neutral. 
This holds true also for other states of the affective spectrum, such as opinions, which are traditionally investigated by research in sentiment analysis. 

Sentiment analysis is a consolidated research field, and a plethora of tools are nowadays publicly available for research purposes. In recent years, a trend emerged and consolidated to leverage sentiment analysis as a tool for empirical software engineering. Recent studies applied sentiment analysis to Stack Overflow, in order to define empirically-driven guidelines for successful question-writing in technical question and answering sites~\cite{CalefatoIST18}. Users' sentiment in app reviews~\cite{Maalej17} or social media~\cite{Guzman16} was studied to support requirements elicitation. Developers' emotions were studied in the context of issue tracking to investigate their impact on issue-fixing time~\cite{Murgia2014} or to understand how emotions are communicated in collaborative-software development~\cite{Gachechiladze17}. Opinion mining on Stack Overflow was leveraged to support the development of recommender systems for software libraries~\cite{Lin2019,Uddin2017}.  

General-purpose sentiment analysis tools have been trained on movie reviews~\cite{Socher2013} or social media texts~\cite{Thelwall2010}. In spite of their popularity, there is a general consensus in the research community about the negative results obtained when using such tools in SE~\cite{Jongeling17,Novielli15}. In our previous work~\cite{Novielli15}, we manually investigated a dataset of 800 posts from Stack Overflow, reporting domain-specific lexicon as the main cause for false positives in negative sentiment detection.  Jongeling and colleagues~\cite{Jongeling17} compared the predictions of widely used off-the-shelf sentiment analysis tools, showing not only how these tools disagree with human annotation of developers’ emotions and opinions, but also how they disagree with each other. They conclude advocating in favor of SE-tuning of sentiment analysis tool to allow reliable empirical studies. 

To overcome these limitations, researchers recently started to develop their own SE-specific tools~\cite{Panichella15, Chen:fse2019,Blaz16,Uddin2017,Lin2019}. At the time of this study, we identified four of the most widely used, SE-specific tools available for research use (see Section~\ref{sec:tools}).  
By replicating the original study of Jongeling et al., our benchmark study \cite{Novielli18} presented at MSR 2018 investigated to what extent fine-tuning sentiment analysis tools for the software engineering domain do succeed in improving the accuracy of emotion detection. The results show that fine-tuning of tools on SE-related text does improve the performance of sentiment classification, provided that the train set used for retraining are built following guidelines grounding on theoretical models of affect. The study was performed in a within-platform setting, that is the train and test sets used for assessing the performance of classifiers based on machine-learning were collected on the same collaborative development platform. In the current study, we partially replicate the experimental setting of the two previous studies~\cite{Jongeling17,Novielli18}. The goal is to further advance the state of the art by addressing the problem of using SE-specific tools in a cross-platform setting, that is when a gold standard dataset is not available for retraining. 

\section{SE-specific Sentiment Analysis Tools}
\label{sec:tools}
State-of-the-art approaches to sentiment analysis treat subjectivity and polarity detection as text classification problems. The existing tools implement two main approaches. The first one exploits \textit{machine learning }algorithms for training supervised classifiers based on textual features. Such features are typically based on words occurring in the documents (i.e., tokens, stems, lemmata) or syntactic features as part-of-speech tags. Often, textual features are extracted using n-grams, i.e., sequences of \textit{n} contiguous words~\cite{Riloff2006}. Such approaches mainly rely on state-of-the-art machine learning algorithms. Recently, researchers also leveraged deep learning~\cite{Zhang2018} in combination with emoji-based vector representations of documents~\cite{Chen:fse2019}. 
On the other hand, \textit{lexicon-based} methods~\cite{Taboada2011} exploit the \textit{prior} sentiment polarity of words in a text, that is the word positive, negative, or neutral polarity based on large lexicons of words annotated with scores indicating their positive or negative semantic orientation. The overall sentiment of a text is then computed based on the prior polarity of the words occurring in it. 
However, due to the effect of contextual valence shifters, such as intensifiers (e.g., adverbs as "very") or negations (e.g., "not"), the prior polarity of a given word might not match the actual sentiment of the author. Therefore, lexicon-based approaches are usually integrated with additional rules to adjust the prior polarity of words based on the effect of intensifiers and negations.

In this study, we assess the performance of four tools that are publicly available at the time of writing. Specifically, Senti4SD and SentiCR leverage supervised machine-learning, while SentiStrength-SE and DEVA implement a lexicon-based approach.

\textbf{Senti4SD}~\cite{CalefatoEMSE18} is our own supervised polarity classifier, which leverages a suite of features based on Bag of Words (BoW), sentiment lexicons, and semantic features based on word embedding.  Along with the toolkit, we distribute a classification model, trained and validated on a gold standard of about 4K questions, answers, and comments from Stack Overflow, and manually annotated for sentiment polarity. Furthermore, the toolkit provides a training method that enables the customization of the classifier using a gold standard as input. Compared to the performance obtained by SentiStrength on the same Stack Overflow test set, Senti4SD reduces the misclassifications of neutral and positive posts as emotionally negative (F1=.87). A good performance (F1=.84) is also achieved with a minimal set of training documents. For this study, we use the Python version of Senti4SD~\cite{EmtkSemotion2019}.

\textbf{SentiCR}~\cite{Ahmed17} is a supervised tool that leverages a feature vector generated by computing term frequency–inverse document frequency (tf-idf) for Bag-of-Words (BoW) extracted from the input text. SentiCR implements basic preprocessing of the raw input text to expand contractions, handle negations and emoticons, remove stop-words, derive word stems, and remove code snippets. Furthermore, it performs SMOTE to handle the class imbalance in the training set. The currently distributed version implements a training approach based on Gradient Boosting Tree and requires a training set as an input in order to retrain the model and use it on the target document collection. A mean accuracy of 83\%, a precision of .68, and a recall of .58 are reported on a gold standard of 2,000 code-review comments. 

\textbf{SentiStrength-SE}~\cite{Islam17} is built upon the API of SentiStrength~\cite{Thelwall2010}. It leverages a manually adjusted version of the SentiStrength lexicon and implements \textit{ad hoc} heuristics to correct the misclassifications observed when running it on a subset of the dataset of Ortu et al.~\cite{Ortu16}. The sentiment scores of words in the lexicon were manually adjusted to reflect the semantics and neutral polarity of domain words such as "support" or "default." As a result, SentiStrength-SE outperforms SentiStrength on technical texts.

\textbf{DEVA}~\cite{Islam18} leverages a lexicon-based approach for the identification of both emotion activation (arousal) and polarity from text. To this end, the tool uses two separate dictionaries developed by exploiting a general-purpose lexicon as well as one specific for software engineering text. To further increase its accuracy, DEVA also includes several heuristics, some of which are borrowed from SentiStrength-SE. For the empirical evaluation, a ground-truth dataset was built, consisting of 1,795 Jira issue comments, manually annotated by three human raters, on which DEVA was found to achieve a precision of .82 and a recall of .79.


\section{Annotated Datasets}
\label{sec:datasets}
The quality of the gold standard largely impacts the classification performance, regardless of the machine learning approach~\cite{Agrawal18,Tantithamthavorn15}. As for sentiment analysis, we found that SE-specific customization might not guarantee a reasonable accuracy if \textit{ad hoc} annotation is performed~\cite{Novielli18}. In fact, \textit{ad hoc} annotation consists of asking the raters to provide polarity labels according to their subjective perception of the semantic orientation of the text ~\cite{Ahmed17,Lin18}. In our previous benchmarking study~\cite{Novielli18}, we provide evidence that the absence of clear guidelines for annotation leads to noisy gold standards, thus resulting in unreliable model training and testing. As such, we argue that reliable sentiment analysis in software engineering is nonetheless possible, provided that manual annotation of gold standards is supported by theoretical models of emotion. In line with our previous findings, in this study, we employ two model-driven datasets from Stack Overflow and Jira (described next), consistently annotated according to the same theoretical framework ~\cite{Shaver87}. As a third dataset, we manually labeled over 7K comments from GitHub pull requests and commits, following the same annotation schema and guidelines used in \cite{CalefatoEMSE18} as detailed next. 
In Table~\ref{table:datasets}, we report the overall number of documents\footnote{In the remainder of the paper, we will use the term `document’ to refer to the text items (posts or comments) in our datasets.} included in each dataset with the distribution of labels for each polarity class. 

\begin{table}[tb]
\caption{Datasets included in our benchmark, with distribution of polarity classes.}
\resizebox{\columnwidth}{!}{%
\begin{tabular}{p{1.6cm}crrr}
\multicolumn{1}{c}{\multirow{2}{*}{\textbf{Dataset}}} & 
\multicolumn{1}{c}{\textbf{Overall}} & \multicolumn{3}{c}{\textbf{Polarity Classes}}\\
\multicolumn{1}{c}{}& {\textbf{documents}} & \textit{Neutral} & \textit{Positive} & \textit{Negative} \\
\hline\\
GitHub & 7,122 & 3,022 (43\%) & 2,013 (28\%) & 2,087 (29\%) \\
Jira~\cite{Ortu16} & 5,869 & 3,955 (67\%) & 1,128 (19\%) & 786 (14\%)
\\Stack Overflow~\cite{CalefatoEMSE18} & 4,423 & 1,694 (38\%) & 1,527 (35\%) & 1,202 (27\%)\\
\end{tabular}
\label{table:datasets}
}
\end{table}

The \textbf{Stack Overflow dataset}~\cite{CalefatoEMSE18} consists of 4,423 posts, including questions, answers, and comments manually annotated with polarity labels by twelve trained coders with a background in Computer Science. The coders were trained to explicitly indicate a polarity label for each post according to the emotion detected, based on the labels included in the Shaver framework~\cite{Shaver87}. Each post was annotated by three raters and received the polarity gold label based on majority voting. The gold standard resulting from this procedure is well-balanced, with 35\% of posts conveying positive emotions, 27\% presenting negative emotions, and 38\% of posts labeled as neutral, denoting the absence of emotions. A Cohen's~\cite{Cohen68} $\kappa$ of .74 is observed, indicating a substantial inter-rater agreement~\cite{Viera05}.

The \textbf{Jira dataset}~\cite{Ortu16} includes about 6,000 issue comments and sentences authored by software developers of popular open-source software projects, such as Apache and Spring. The Jira dataset is originally distributed with the six emotion labels from the Shaver et al. framework~\cite{Shaver87} (i.e., love, joy, surprise, anger, fear, and sadness), whereas this study focuses on emotion polarity (i.e., the positive, negative, or neutral valence conveyed by texts). As such, we use an approach consistent with the labeling guidelines adopted for the Stack Overflow gold standard described above, thus resulting in two homogeneous benchmarking datasets grounded on the same emotion model. Specifically, we translate positive emotions, i.e., love and joy, into a positive polarity label. Similarly, sadness, anger, and fear are mapped to the negative polarity class. Instead, surprise cases are discarded as this emotion label could be either considered positive or negative, depending on the expectations of the author of a text. Finally, the absence of emotions defines neutral cases. Unlike the Stack Overflow dataset, the Jira gold standard is not well-balanced, with 19\% of posts conveying positive emotions, 14\% conveying negative emotions, and 67\% labeled as neutral. The authors do not assess the $\kappa$ agreement for the polarity classes, as they originally provide labels for discrete emotions. Conversely, they report the $\kappa$ for the emotion annotation, with values ranging from absence of agreement for \textit{anger} to moderate agreement for \textit{love}, for which the highest value observed is $\kappa = .55$.

The \textbf{GitHub dataset} includes about 7,000 pull request and commit comments. The dataset is well-balanced, is a desirable property for a training set~\cite{He2009}, with 28\% and 29\% of posts conveying positive and negative emotions, respectively. The remaining 43\% of posts are labeled as neutral, as they do not convey emotions. The dataset has been annotated by three of the authors following the guidelines for annotation defined for the creation of the Stack Overflow dataset~\cite{CalefatoEMSE18}. As a unit of analysis, we consider the entire comment, i.e., the raters were requested to annotate the sentiment conveyed by the whole comment. Specifically, the raters were trained to provide a polarity label based on the emotion detected according to the Shaver model, following the emotion-polarity mapping described for the Stack Overflow and Jira datasets. 

\begin{figure}[h]
\includegraphics[width=\columnwidth]{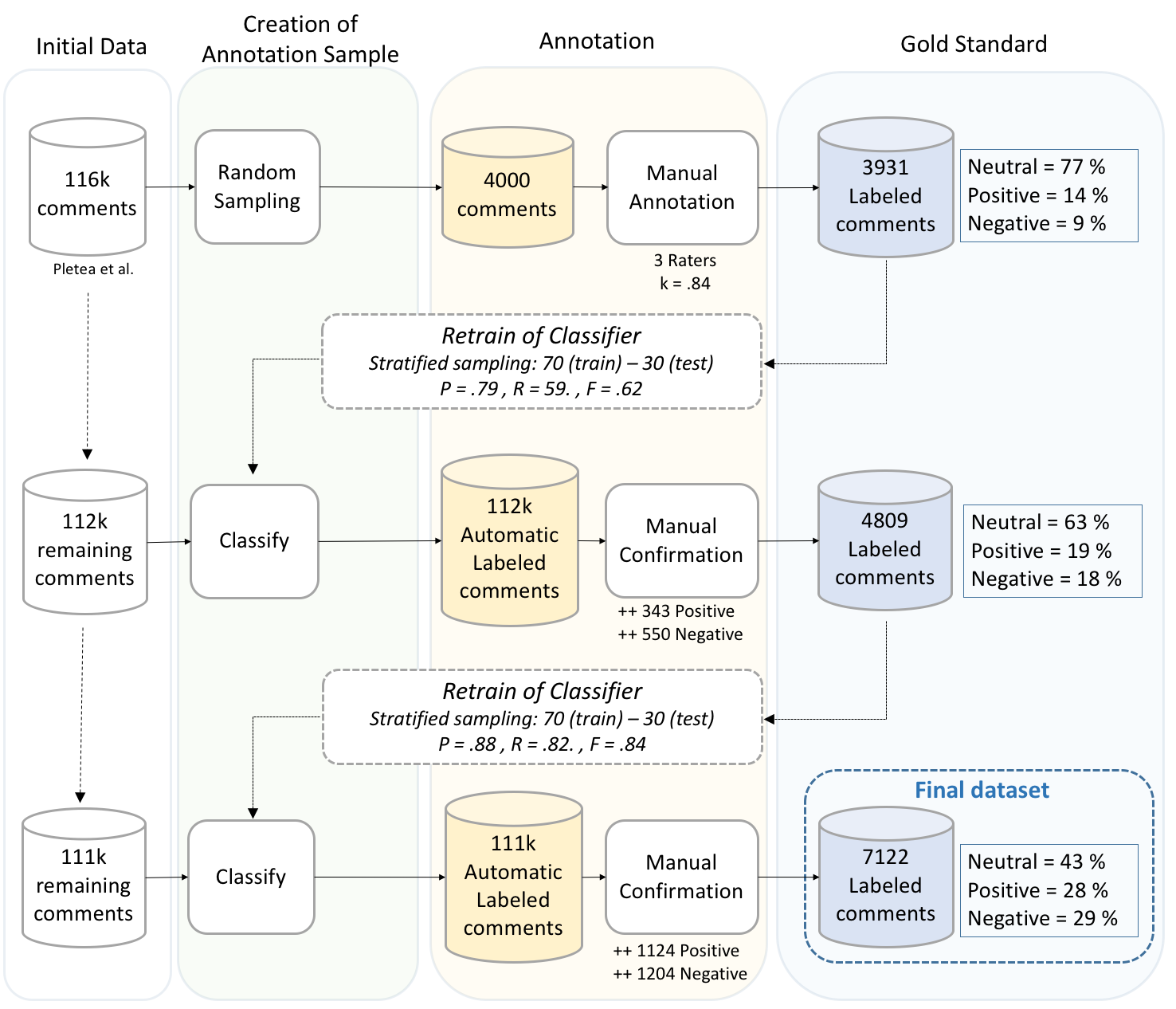}
\caption{Creating the Gold Standard through manual annotation of polarity classes.}
\label{fig:goldstandard}
\end{figure}

We built our GitHub gold standard using the iterative approach depicted in Figure~\ref{fig:goldstandard}. Specifically, we designed the protocol for our annotation study following the methodology adopted in the study on anger in collaborative software development ~\cite{Gachechiladze17}. We extracted the annotation sample for each iteration from the dataset of comments created to study the sentiment of security discussion in GitHub ~\cite{Pletea14}. We started with an annotation sample of 4k comments, randomly extracted from the initial dataset of 116k comments. Each comment was labeled by two raters independently. We observed an almost perfect inter-rater agreement ($\kappa$ = .84). Once the individual annotation was completed, we assigned the manually provided gold label to all the comments for which the two raters agreed. Then, the three raters discussed the 340 disagreement cases in a plenary meeting: we include in the gold standard all those comments for which the initial disagreement is resolved through discussion (298) and discard the others (42, corresponding to 1.05\% of the annotation sample). Furthermore, 27 duplicate comments were removed.    

As a result of this first step, we obtained 3,931 comments for which the three raters agreed both on the presence of emotions and on its polarity. Given the unbalanced distribution of the obtained dataset (see Figure~\ref{fig:goldstandard}), we implemented the subsequent two annotation steps to collect more positive and negative comments. Since manual labeling is a time-consuming activity, we accelerated the process by leveraging a semi-automatic approach involving manual confirmation of automatically obtained polarity labels. Using the initial core of 3,931 comments, we retrained the polarity classification model using the Senti4SD toolkit, as it reported a better precision than SentiCR for both the positive and negative classes. Specifically, we observe a precision of .61 (Senti4SD) vs. .34 (SentiCR) for the negative class. Conversely, the precision for the positive class is comparable (.89 for Senti4SD vs. 88 for SentiCR). The reason behind this choice is that, by optimizing for precision, we reduce the number of neutral sentences misclassified as expressing sentiment, thus avoiding to annoy the raters with useless annotation of neutral cases. The performance of this classification model is reported in Figure~\ref{fig:goldstandard} (Precision = .79, Recall =.59, F1-measure = .62).  

In the second step, we applied this classifier to the remaining 112k comments of the original dataset by Pletea et al., excluding all cases that were already included in the first annotation sample. We obtained an automatically labeled dataset, from which we randomly extracted a new annotation sample of 600 positive and 600 negative comments. To avoid any bias, the annotators were not provided with the outcome of the classifier. As such, their annotation was done only based on the text, as in the first round. Again, the raters performed the annotation individually. They confirmed the classifier label for 343 positive and 550 negative comments. These new confirmed cases were added to the gold standard, resulting in an enriched set of 4,809 comments, of which 63\% labeled as neutral, 19\% as positive, and 18\% as negative. To further enrich and balance the gold standard, we repeated the training with this new set, observing an improved performance of the classification model (Precision = .88, Recall =.82, F1-measure = .84). We use this second classifier to label the remaining 111k comments and repeat the manual confirmation step for 3,000 comments. This third annotation step resulted in 1,124 positive and 1,204  additional negative comments. The final GitHub gold standard includes 7,122 comments that we use for this study.

\section{Study Design}
\label{sec:experimental-setting}
\textbf{Experimental Setting.} To answer RQ1 and RQ2, we assess the performance on the three gold standard datasets of the two supervised tools (Senti4SD and SentiCR), which can be retrained, and the two lexicon-based classifiers (SentiStrength-SE and DEVA), for which retraining is not possible. To enable comparison with the within-platform benchmark ~\cite{Novielli18}, we replicate the former experimental setting. Specifically, we split each gold set into training (70\%) and test (30\%) sets by performing stratified sampling with scikit-learn.\footnote{\url{https://scikit-learn.org/stable/index.html}} We evaluate the performance of all tools on the held-out test sets. As for the supervised classifiers, we first use the training set to retrain them using the methods provided by each toolkit; then, their performance is assessed in a cross-platform setting, by using the test set from the other experimental datasets (e.g., we train on the 70\% train set of Stack Overflow and test on the 30\% test sets from Jira and GitHub). Furthermore, we run twice the train and test for Senti4SD, because the feature set of Senti4SD can be customized. As such, we also run the train/test steps by removing the keyword-based features, that is, the uni- and bi-grams Bag of Words (BoW). The reason behind this choice is to understand the extent to which the interaction style, i.e., the specific lexicon or jargon observed in a given platform, has an impact on the performance. We cannot replicate this evaluation for SentiCR as it only exploits features based on BoW.

In this study, we aim to compare the results we achieved when training/testing supervised approaches on different datasets (cross-platform) with the performance observed when such approaches are trained and tested on the same dataset (within-platform), as done in our previous work ~\cite{Novielli18}. 
However, even minor changes in the settings used in the two studies may lead to major differences in the results. To address this problem, we rerun the training/testing in the within-platform setting for comparison, following the approach we previously adopted and described in~\cite{Novielli18} and observed negligible differences in the tool performances. By doing so, all possible confounding factors are controlled, and we can be confident that the potential differences between the two scenarios (within- vs. cross-platform settings) would be due to the used training sets. 

To address RQ3, we analyze the learning curves of the supervised tools in a within-platform setting. The goal of this evaluation is to identify the minimum size of the gold standard that makes re-training convenient for supervised tools as compared to using lexicon-based, non-customizable ones. The learning curves enable us to visually assess how the size of the training set influences the classification performance. We start by training the supervised tools using a subset of 5\% of the original training set for each platform. At each step, we increment the training set size by 5\%. At every iteration, the subset for training is extracted from scratch with stratified sampling, and the performance is assessed on the entire held-out 30\% test set from the same platform. Given the unbalanced distribution of the Jira dataset, we repeated the performance of evaluation twice for Senti4SD, with and without performing data resampling. To enable comparison with SentiCR, we use SMOTE~\cite{Chawla02} by replicating the setting described by its authors~\cite{Ahmed17}. 

\textbf{Metrics}. We report the performance of each sentiment analysis tool in terms of precision, recall, and F1-measure for all the three polarity classes. This choice is in line with previous research~\cite{Jongeling17,Novielli18} and is consistent with the standard methodology adopted for benchmarking of sentiment analysis systems as well as more general text categorization approaches in evaluations campaigns~\cite{Caselli18}. 

For the sake of completeness, we report the overall performance using both micro- and macro-averaging as aggregated metrics. Micro-averaging is known to be influenced by the performance on the majority class~\cite{Sebastiani02}. Conversely, the ability of a classifier to correctly identify items belonging to classes with few training instances is correctly assessed by the macro-average. Given the unbalanced distribution of the Jira dataset, in this study we rely on the macro-average, i.e., precision and recall are first evaluated locally for each class, and then globally by averaging the results of the different categories. 

Furthermore, we use the weighted kappa ($\kappa$)~\cite{Cohen68,Viera05} to assess both the agreement with gold labels (RQ1) and the agreement among the three tools (RQ2). 
We distinguish between mild disagreement (weight = 1), i.e., the disagreement between negative/positive and neutral annotations, and strong disagreement (weight = 2), i.e., the disagreement between positive and negative judgments. We interpret $\kappa$ as follows ~\cite{Viera05}: $\kappa$ values less or equal to zero indicate that agreement is less than chance; the agreement is slight if $0.01 \leq\kappa\leq 0.20$, fair if $0.21 \leq\kappa\leq 0.40$, moderate if $0.41 \leq\kappa\leq 0.60$, substantial if $0.61 \leq\kappa\leq 0.80$ and almost perfect if $0.81 \leq\kappa\leq 1$. Both the weighted scheme and the interpretation of $\kappa$ are the same adopted in the previous studies~\cite{Jongeling17, Novielli18}.

\section{Results}
\label{sec:results}

\subsection{Performance of SE-specific tools in cross-platform settings}
\label{sec:performance}
\paragraph{\textbf{RQ1} - To what extent do different SE-specific sentiment analysis tools agree with the emotions of software developers when used as ‘off-the-shelf’ tools in a cross-platform setting?}   

\begin{table*}[h!]
\caption{Performance of SE-specific sentiment analysis tools in the cross-platform setting. For each setting, we highlight the best values for each metric in bold and the overall performance in Italic. The within-platform setting is reported in grey.}
\label{tab:results}
\resizebox{\textwidth}{!}{%
\begin{tabular}{ccl|lll|lll|lll|lll|lll}
\hline
\multirow{2}{*}{\textbf{Setting}} & \multirow{2}{*}{\textbf{Train set}} & \multicolumn{1}{c}{\multirow{2}{*}{\textbf{\begin{tabular}[c]{@{}c@{}}Polarity\\ Class\end{tabular}}}} & \multicolumn{3}{|c|}{\textbf{Senti4SD}} & \multicolumn{3}{|c|}{\textbf{Senti4SD (no BoW)}} & \multicolumn{3}{|c|}{\textbf{SentiCR}} & \multicolumn{3}{|c|}{\textbf{SentiStrength-SE}} & \multicolumn{3}{|c|}{\textbf{DEVA}} \\ \cline{4-18} 
 &  & \multicolumn{1}{c}{} & \multicolumn{1}{c}{\textit{P}} & \multicolumn{1}{c}{\textit{R}} & \multicolumn{1}{c}{\textit{F1}} & \multicolumn{1}{c}{\textit{P}} & \multicolumn{1}{c}{\textit{R}} & \multicolumn{1}{c}{\textit{F1}} & \multicolumn{1}{c}{\textit{P}} & \multicolumn{1}{c}{\textit{R}} & \multicolumn{1}{c}{\textit{F1}} & \multicolumn{1}{c}{\textit{P}} & \multicolumn{1}{c}{\textit{R}} & \multicolumn{1}{c}{\textit{F1}} & \multicolumn{1}{c}{\textit{P}} & \multicolumn{1}{c}{\textit{R}} & \multicolumn{1}{c}{\textit{F1}} \\ \hline
\multicolumn{18}{|c|}{\textit{Test set: GitHub}} \\ \hline
\multirow{5}{*}{Within-platform} & \multirow{5}{*}{GitHub} & \cellcolor[HTML]{EFEFEF}Negative &\cellcolor[HTML]{EFEFEF}.92 &\cellcolor[HTML]{EFEFEF} .90 &\cellcolor[HTML]{EFEFEF} .91 & \multicolumn{3}{|c|}{\multirow{5}{*}{\cellcolor[HTML]{EFEFEF}--}} &\cellcolor[HTML]{EFEFEF} .90 &\cellcolor[HTML]{EFEFEF} .63 &\cellcolor[HTML]{EFEFEF} .74 &\cellcolor[HTML]{EFEFEF} .79 &\cellcolor[HTML]{EFEFEF} .77 &\cellcolor[HTML]{EFEFEF} .78 &\cellcolor[HTML]{EFEFEF} .65 &\cellcolor[HTML]{EFEFEF} .68 &\cellcolor[HTML]{EFEFEF} .67 \\
 &  & \cellcolor[HTML]{EFEFEF}Neutral & \cellcolor[HTML]{EFEFEF}.90 & \cellcolor[HTML]{EFEFEF}.93 & \cellcolor[HTML]{EFEFEF}.92 & \multicolumn{3}{|c|}{\cellcolor[HTML]{EFEFEF}} &\cellcolor[HTML]{EFEFEF}.76 &\cellcolor[HTML]{EFEFEF} .94 &\cellcolor[HTML]{EFEFEF} .84 &\cellcolor[HTML]{EFEFEF} .78 &\cellcolor[HTML]{EFEFEF} .86 &\cellcolor[HTML]{EFEFEF} .82 &\cellcolor[HTML]{EFEFEF} .83 &\cellcolor[HTML]{EFEFEF} .71 &\cellcolor[HTML]{EFEFEF} .77 \\
 &  & \cellcolor[HTML]{EFEFEF}Positive &\cellcolor[HTML]{EFEFEF}.95 &\cellcolor[HTML]{EFEFEF} .91 &\cellcolor[HTML]{EFEFEF} .93 &\multicolumn{3}{|c|}{\cellcolor[HTML]{EFEFEF} --} &\cellcolor[HTML]{EFEFEF} .89 &\cellcolor[HTML]{EFEFEF} .85 &\cellcolor[HTML]{EFEFEF} .87 &\cellcolor[HTML]{EFEFEF}.86 & \cellcolor[HTML]{EFEFEF}.76 &\cellcolor[HTML]{EFEFEF} .81 &\cellcolor[HTML]{EFEFEF} .69 &\cellcolor[HTML]{EFEFEF} .81 &\cellcolor[HTML]{EFEFEF} .75 \\
 &  & \cellcolor[HTML]{EFEFEF}\textit{Micro-avg.} & \cellcolor[HTML]{EFEFEF}\textit{.92} & \cellcolor[HTML]{EFEFEF}\textit{.92} & \cellcolor[HTML]{EFEFEF}\textit{.92} & \multicolumn{3}{|c|}{\cellcolor[HTML]{EFEFEF} } & \cellcolor[HTML]{EFEFEF}\textit{.82} & \cellcolor[HTML]{EFEFEF}\textit{.82} & \cellcolor[HTML]{EFEFEF}\textit{.82} & \cellcolor[HTML]{EFEFEF}\textit{.80} & \cellcolor[HTML]{EFEFEF}\textit{.80} & \cellcolor[HTML]{EFEFEF}\textit{.80} & \cellcolor[HTML]{EFEFEF}\textit{.73} & \cellcolor[HTML]{EFEFEF}\textit{.73} & \cellcolor[HTML]{EFEFEF}\textit{.73}\\
 &  & \cellcolor[HTML]{EFEFEF}\textit{Macro-avg.} & \cellcolor[HTML]{EFEFEF}\textit{.92} & \cellcolor[HTML]{EFEFEF}\textit{.92} & \cellcolor[HTML]{EFEFEF}\textit{.92} & \multicolumn{3}{|c|}{\cellcolor[HTML]{EFEFEF}} & \cellcolor[HTML]{EFEFEF}\textit{.85} & \cellcolor[HTML]{EFEFEF}\textit{.81} & \cellcolor[HTML]{EFEFEF}\textit{.82} & \cellcolor[HTML]{EFEFEF}\textit{.81} & \cellcolor[HTML]{EFEFEF}\textit{.80} & \cellcolor[HTML]{EFEFEF}\textit{.80} & \cellcolor[HTML]{EFEFEF}\textit{.72} & \cellcolor[HTML]{EFEFEF}\textit{.73} & \cellcolor[HTML]{EFEFEF}\textit{.73} \\
\hline
\multirow{5}{*}{Cross-platform} & \multirow{5}{*}{Stack Overflow} & Negative & \textbf{.79} & .50 & .61 & .75 & \textbf{.83} & \textbf{.79} & .78 & .34 & .47 & .79 & .77 & .78 & .65 & .68 & .67 \\
 &  & Neutral & .71 & .85 & .77 & .82 & .78 & .80 & .60 & \textbf{.93} & .73 & .78 & .86 & \textbf{.82} & \textbf{.83} & .71 & .77 \\
 &  & Positive & .76 & .84 & .80 & \textbf{.88} & \textbf{.85} & \textbf{.86} & .86 & .67 & .75 & .86 & .76 & .81 & .69 & .81 & .75 \\
 &  & \textit{Micro-avg.} & \textit{.74} & \textit{.74} & \textit{.74} & \textit{\textbf{.82}} & \textit{\textbf{.82}} & \textit{\textbf{.82}} & \textit{.68} & \textit{.68} & \textit{.68} & \textit{.80} & \textit{.80} & \textit{.80} & \textit{.73} & \textit{.73} & \textit{.73} \\
 &  & \textit{Macro-avg.} & \textit{.76} & \textit{\textbf{.84}} & \textit{.80} & \textit{\textbf{.82}} & \textit{.82} & \textit{\textbf{.82}} & \textit{.75} & \textit{.65} & \textit{.65} & \textit{.81} & \textit{.80} & \textit{.80} & \textit{.72} & \textit{.73} & \textit{.73} \\
\multicolumn{2}{r}{\multirow{2}{*}{\textit{Differences with the within-platform setting}}} & \textit{Micro-avg.} & \textit{-.18} & \textit{-.18} & \textit{-.18} & \textit{-.10} & \textit{-.10} & \textit{-.10} & \textit{-.14} & \textit{-.14} & \textit{-.14} & \multicolumn{3}{|c|}{\multirow{2}{*}{--}} & \multicolumn{3}{|c}{\multirow{2}{*}{--}} \\
\multicolumn{2}{r}{} & \textit{Macro-avg.} & \textit{-.16} & \textit{-.08} & \textit{-.12} & \textit{-.10} & \textit{-.10} & \textit{-.10} & \textit{-.10} & \textit{-.16} & \textit{-.17} & \multicolumn{3}{|c|}{} & \multicolumn{3}{|c}{} \\
\hline
\multirow{5}{*}{Cross-platform} & \multirow{5}{*}{Jira} & Negative & .84 & .51 & .63 & \textbf{.86} & .50 & .64 & .84 & .24 & .37 & .79 & \textbf{.77} & \textbf{.78} & .65 & .68 & .67 \\
 &  & Neutral & .59 & .96 & .73 & .62 & .95 & .75 & .51 & \textbf{.98} & .67 & .78 & .86 & \textbf{.82} & \textbf{.83} & .71 & .77 \\
 &  & Positive & \textbf{.93} & .45 & .61 & .91 & .57 & .70 & .92 & .35 & .51 & .86 & .76 & \textbf{.81} & .69 & \textbf{.81} & .75 \\
 &  & \textit{Micro-avg.} & \textit{.68} & \textit{.68} & \textit{.68} & \textit{.71} & \textit{.71} & \textit{.71} & \textit{.58} & \textit{.58} & \textit{.58} & \textit{\textbf{.80}} & \textit{\textbf{.80}} & \textit{\textbf{.80}} & \textit{.73} & \textit{.73} & \textit{.73} \\
 &  & \textit{Macro-avg.} & \textit{.79} & \textit{.64} & \textit{.66} & \textit{.79} & \textit{.67} & \textit{.69} & \textit{.76} & \textit{.52} & \textit{.52} & \textit{\textbf{.81}} & \textit{\textbf{.80}} & \textit{\textbf{.80}} & \textit{.72} & \textit{.73} & \textit{.73} \\
\multicolumn{2}{r}{\multirow{2}{*}{\textit{Differences with the within-platform setting}}} & \textit{Micro-avg.} & \textit{-.24} & \textit{-.24} & \textit{-.24} & \textit{-.21} & \textit{-.21} & \textit{-.21} & \textit{-.24} & \textit{-.24} & \textit{-.24} & \multicolumn{3}{|c}{\multirow{2}{*}{\textit{--}}} & \multicolumn{3}{|c}{\multirow{2}{*}{\textit{--}}} \\
\multicolumn{2}{r}{} & \textit{Macro-avg.} & \textit{-.13} & \textit{-.28} & \textit{-.26} & \textit{-.13} & \textit{-.25} & \textit{-.23} & \textit{-.09} & \textit{-.29} & \textit{-.30} & \multicolumn{3}{|c}{} & \multicolumn{3}{|c}{} \\ \hline
\multicolumn{18}{|c|}{\textit{Test set: Stack Overflow}} \\ \hline
\multirow{5}{*}{Within-platform} & \multirow{5}{*}{Stack Overflow} &\cellcolor[HTML]{EFEFEF} Negative &\cellcolor[HTML]{EFEFEF} .81 &\cellcolor[HTML]{EFEFEF} .87 &\cellcolor[HTML]{EFEFEF} .84 &\multicolumn{3}{|c|}{\multirow{5}{*}{\cellcolor[HTML]{EFEFEF} --}} &\cellcolor[HTML]{EFEFEF} .79 &\cellcolor[HTML]{EFEFEF} .74 &\cellcolor[HTML]{EFEFEF} .76 &\cellcolor[HTML]{EFEFEF}.74 &\cellcolor[HTML]{EFEFEF} .79 &\cellcolor[HTML]{EFEFEF} .76 &\cellcolor[HTML]{EFEFEF} .67 &\cellcolor[HTML]{EFEFEF} .79 &\cellcolor[HTML]{EFEFEF} .73 \\
 &  &\cellcolor[HTML]{EFEFEF} Neutral &\cellcolor[HTML]{EFEFEF} .86 &\cellcolor[HTML]{EFEFEF} .81 &\cellcolor[HTML]{EFEFEF} .84 & \multicolumn{3}{|c|}{\cellcolor[HTML]{EFEFEF} } &\cellcolor[HTML]{EFEFEF} .80 &\cellcolor[HTML]{EFEFEF} .82 &\cellcolor[HTML]{EFEFEF} .81 &\cellcolor[HTML]{EFEFEF} .76 &\cellcolor[HTML]{EFEFEF} .76 &\cellcolor[HTML]{EFEFEF} .76 &\cellcolor[HTML]{EFEFEF} .84 &\cellcolor[HTML]{EFEFEF} .68 &\cellcolor[HTML]{EFEFEF} .75 \\
 &  &\cellcolor[HTML]{EFEFEF} Positive &\cellcolor[HTML]{EFEFEF} .91 &\cellcolor[HTML]{EFEFEF} .92 &\cellcolor[HTML]{EFEFEF} .92 & \multicolumn{3}{|c|}{\cellcolor[HTML]{EFEFEF} --} &\cellcolor[HTML]{EFEFEF} .88 &\cellcolor[HTML]{EFEFEF} .90 &\cellcolor[HTML]{EFEFEF} .89 &\cellcolor[HTML]{EFEFEF} .89 &\cellcolor[HTML]{EFEFEF} .84 &\cellcolor[HTML]{EFEFEF} .86 &\cellcolor[HTML]{EFEFEF} .85 &\cellcolor[HTML]{EFEFEF} .90 &\cellcolor[HTML]{EFEFEF} .87 \\
 &  & \cellcolor[HTML]{EFEFEF}\textit{Micro-avg.} & \cellcolor[HTML]{EFEFEF}\textit{.87} & \cellcolor[HTML]{EFEFEF}\textit{.87} & \cellcolor[HTML]{EFEFEF}\textit{.87} & \multicolumn{3}{|c|}{\cellcolor[HTML]{EFEFEF} } & \cellcolor[HTML]{EFEFEF}\textit{.83} & \cellcolor[HTML]{EFEFEF}\textit{.83} & \cellcolor[HTML]{EFEFEF}\textit{.83} & \cellcolor[HTML]{EFEFEF}\textit{.80} & \cellcolor[HTML]{EFEFEF}\textit{.80} & \cellcolor[HTML]{EFEFEF}\textit{.80} &\cellcolor[HTML]{EFEFEF}\textit{.79} & \cellcolor[HTML]{EFEFEF}\textit{.79} & \cellcolor[HTML]{EFEFEF}\textit{.79} \\
 &  & \cellcolor[HTML]{EFEFEF}\textit{Macro-avg.} & \cellcolor[HTML]{EFEFEF}\textit{.86} & \cellcolor[HTML]{EFEFEF}\textit{.87} & \cellcolor[HTML]{EFEFEF}\textit{.87} & \multicolumn{3}{|c|}{\cellcolor[HTML]{EFEFEF} } & \cellcolor[HTML]{EFEFEF}\textit{.82} & \cellcolor[HTML]{EFEFEF}\textit{.82} & \cellcolor[HTML]{EFEFEF}\textit{.82} & \cellcolor[HTML]{EFEFEF}\textit{.80} & \cellcolor[HTML]{EFEFEF}\textit{.80} & \cellcolor[HTML]{EFEFEF}\textit{.79} & \cellcolor[HTML]{EFEFEF}\textit{.79} & \cellcolor[HTML]{EFEFEF}\textit{.79} & \cellcolor[HTML]{EFEFEF}\textit{.78} \\
\hline
\multirow{5}{*}{Cross-platform} & \multirow{5}{*}{GitHub} & Negative & .71 & .72 & .72 & .71 & \textbf{.80} & .75 & .72 & .41 & .52 & \textbf{.74} & .79 & \textbf{.76} & .67 & .79 & .73 \\
 &  & Neutral & .71 & .80 & .75 & .79 & .75 & \textbf{.77} & .64 & \textbf{.88} & .74 & .76 & .76 & .76 & \textbf{.84} & .68 & .75 \\
 &  & Positive & \textbf{.92} & .78 & .84 & .90 & .85 & \textbf{.87} & .84 & .78 & .81 & .89 & .84 & .86 & .85 & \textbf{.90} & \textbf{.87} \\
 &  & \textit{Micro-avg.} & \textit{.77} & \textit{.77} & \textit{.77} & \textit{\textbf{.80}} & \textit{\textbf{.80}} & \textit{\textbf{.80}} & \textit{.72} & \textit{.72} & \textit{.72} & \textit{\textbf{.80}} & \textit{\textbf{.80}} & \textit{\textbf{.80}} & \textit{.79} & \textit{.79} & \textit{.79} \\
 &  & \textit{Macro-avg.} & \textit{.78} & \textit{.77} & \textit{.77} & \textit{\textbf{.80}} & \textit{\textbf{.80}} & \textit{\textbf{.80}} & \textit{.73} & \textit{.69} & \textit{.69} & \textit{\textbf{.80}} & \textit{\textbf{.80}} & \textit{.79} & \textit{.79} & \textit{.79} & \textit{.78} \\
\multicolumn{2}{r}{\multirow{2}{*}{\textit{Differences with the within-platform setting}}} & \textit{Micro-avg.} & \textit{-.10} & \textit{-.10} & \textit{-.10} & \textit{-.07} & \textit{-.07} & \textit{-.07} & \textit{-.11} & \textit{-.11} & \textit{-.11} & \multicolumn{3}{|c}{\multirow{2}{*}{\textit{--}}} & \multicolumn{3}{|c}{\multirow{2}{*}{\textit{--}}} \\
\multicolumn{2}{r}{} & \textit{Macro-avg.} & \textit{-.09} & \textit{-.10} & \textit{-.10} & \textit{-.06} & \textit{-.07} & \textit{-.07} & \textit{-.09} & \textit{-.13} & \textit{-.13} & \multicolumn{3}{|c|}{} & \multicolumn{3}{|c}{} \\
\hline
\multirow{5}{*}{Cross-platform} & \multirow{5}{*}{Jira} & Negative & .77 & .34 & .48 & \textbf{.79} & .32 & .46 & .71 & .13 & .22 & .74 & \textbf{.79} & \textbf{.76} & .67 & \textbf{.79} & .73 \\
 &  & Neutral & .56 & .93 & .70 & .57 & .93 & .70 & .46 & \textbf{.97} & .62 & .76 & .76 & \textbf{.76} & \textbf{.84} & .68 & .75 \\
 &  & Positive & \textbf{.96} & .65 & .78 & .92 & .70 & .79 & \textbf{.96} & .38 & .54 & .89 & .84 & .86 & .85 & \textbf{.90} & \textbf{.87} \\
 &  & Micro-avg. & .68 & .68 & .68 & .68 & .68 & .68 & .54 & .54 & .54 & \textbf{.80} & .\textbf{80} & \textbf{.80} & \textbf{.79} & .79 & .79 \\
 &  & Macro-avg. & .76 & .64 & .65 & .76 & .65 & .65 & .71 & .49 & .46 & \textbf{.80} & \textbf{.80} & .79 & .79 & .79 & .78 \\
\multicolumn{2}{r}{\multirow{2}{*}{\textit{Differences with the within-platform setting}}} & \textit{Micro-avg.} & \textit{-.19} & \textit{-.19} & \textit{-.19} & \textit{-.19} & \textit{-.19} & \textit{-.19} & \textit{-.29} & \textit{-.29} & \textit{-.29} & \multicolumn{3}{|c}{\multirow{2}{*}{\textit{--}}} & \multicolumn{3}{|c}{\multirow{2}{*}{\textit{--}}} \\
\multicolumn{2}{r}{} & \textit{Macro-avg.} & \textit{-.10} & \textit{-.23} & \textit{-.22} & \textit{-.10} & \textit{-.22} & \textit{-.22} & \textit{-.11} & \textit{-.33} & \textit{-.36} & \multicolumn{3}{|c|}{} & \multicolumn{3}{|c}{} \\ \hline
\multicolumn{18}{|c|}{\textit{Test set: Jira}} \\ \hline
\multirow{5}{*}{Within-platform} & \multirow{5}{*}{Jira} &\cellcolor[HTML]{EFEFEF} Negative &\cellcolor[HTML]{EFEFEF} .75 &\cellcolor[HTML]{EFEFEF} .60 &\cellcolor[HTML]{EFEFEF} .67 & \multicolumn{3}{|c|}{\multirow{5}{*}{\cellcolor[HTML]{EFEFEF} --}} &\cellcolor[HTML]{EFEFEF} .83 &\cellcolor[HTML]{EFEFEF} .63 &\cellcolor[HTML]{EFEFEF} .72 &\cellcolor[HTML]{EFEFEF} .64 &\cellcolor[HTML]{EFEFEF} .72 &\cellcolor[HTML]{EFEFEF} .68 &\cellcolor[HTML]{EFEFEF} .52 &\cellcolor[HTML]{EFEFEF} .69 &\cellcolor[HTML]{EFEFEF} .59 \\
 &  &\cellcolor[HTML]{EFEFEF} Neutral &\cellcolor[HTML]{EFEFEF} .87 &\cellcolor[HTML]{EFEFEF} .89 &\cellcolor[HTML]{EFEFEF} .88 & \multicolumn{3}{|c|}{\cellcolor[HTML]{EFEFEF} } &\cellcolor[HTML]{EFEFEF}.88 &\cellcolor[HTML]{EFEFEF} .91 &\cellcolor[HTML]{EFEFEF} .89 &\cellcolor[HTML]{EFEFEF} .93 & \cellcolor[HTML]{EFEFEF}.81 &\cellcolor[HTML]{EFEFEF} .87 &\cellcolor[HTML]{EFEFEF} .96 &\cellcolor[HTML]{EFEFEF} .75 &\cellcolor[HTML]{EFEFEF} .83 \\
 &  &\cellcolor[HTML]{EFEFEF} Positive &\cellcolor[HTML]{EFEFEF} .76 &\cellcolor[HTML]{EFEFEF} .78 &\cellcolor[HTML]{EFEFEF} .77 & \multicolumn{3}{|c|}{\cellcolor[HTML]{EFEFEF} --} &\cellcolor[HTML]{EFEFEF} .79 &\cellcolor[HTML]{EFEFEF} .81 &\cellcolor[HTML]{EFEFEF} .80 &\cellcolor[HTML]{EFEFEF} .69 & \cellcolor[HTML]{EFEFEF}.93 &\cellcolor[HTML]{EFEFEF} .79 &\cellcolor[HTML]{EFEFEF} .63 &\cellcolor[HTML]{EFEFEF} .90 &\cellcolor[HTML]{EFEFEF} .74 \\
 &  & \cellcolor[HTML]{EFEFEF}\textit{Micro-avg.} & \cellcolor[HTML]{EFEFEF}\textit{.83} & \cellcolor[HTML]{EFEFEF}\textit{.83} & \cellcolor[HTML]{EFEFEF}\textit{.83} & \multicolumn{3}{|c|}{\cellcolor[HTML]{EFEFEF} } & \cellcolor[HTML]{EFEFEF}\textit{.86} & \cellcolor[HTML]{EFEFEF}\textit{.86} & \cellcolor[HTML]{EFEFEF}\textit{.86} & \cellcolor[HTML]{EFEFEF}\textit{.82} & \cellcolor[HTML]{EFEFEF}\textit{.82} & \cellcolor[HTML]{EFEFEF}\textit{.82} & \cellcolor[HTML]{EFEFEF}\textit{.77} & \cellcolor[HTML]{EFEFEF}\textit{.77} & \cellcolor[HTML]{EFEFEF}\textit{.77} \\
 &  & \cellcolor[HTML]{EFEFEF}\textit{Macro-avg.} & \cellcolor[HTML]{EFEFEF}\textit{.79} & \cellcolor[HTML]{EFEFEF}\textit{.76} & \cellcolor[HTML]{EFEFEF}\textit{.77} & \multicolumn{3}{|c|}{\cellcolor[HTML]{EFEFEF} } & \cellcolor[HTML]{EFEFEF}\textit{.83} & \cellcolor[HTML]{EFEFEF}\textit{.78} & \cellcolor[HTML]{EFEFEF}\textit{.80} & \cellcolor[HTML]{EFEFEF}\textit{.75} & \cellcolor[HTML]{EFEFEF}\textit{.82} & \cellcolor[HTML]{EFEFEF}\textit{.78} & \cellcolor[HTML]{EFEFEF}\textit{.69} & \cellcolor[HTML]{EFEFEF}\textit{.78} & \cellcolor[HTML]{EFEFEF}\textit{.72} \\
\hline
\multicolumn{1}{l}{\multirow{5}{*}{Cross-platform}} & \multicolumn{1}{c}{\multirow{5}{*}{GitHub}} & Negative & .57 & .64 & .61 & .57 & .61 & .59 & \textbf{.75} & .56 & .64 & .64 & \textbf{.72} & \textbf{.68} & .52 & .69 & .59 \\
\multicolumn{1}{l}{} & \multicolumn{1}{l}{} & Neutral & .89 & .79 & .84 & .88 & .80 & .84 & .90 & \textbf{.87} & \textbf{.88} & .93 & .81 & .87 & \textbf{.96} & .75 & .83 \\
\multicolumn{1}{l}{} & \multicolumn{1}{l}{} & Positive & .65 & .85 & .74 & .65 & .81 & .72 & \textbf{.71} & .91 & \textbf{.8} & .69 & \textbf{.93} & .79 & .63 & .9 & .74 \\
\multicolumn{1}{l}{} & \multicolumn{1}{l}{} & \textit{Micro-avg.} & \textit{.78} & \textit{.78} & \textit{.78} & \textit{.78} & \textit{.78} & \textit{.78} & \textit{\textbf{.84}} & \textit{\textbf{.84}} & \textit{\textbf{.84}} & \textit{.82} & \textit{.82} & \textit{.82} & \textit{.77} & \textit{.77} & \textit{.77} \\
\multicolumn{1}{l}{} & \multicolumn{1}{l}{} & \textit{Macro-avg.} & \textit{.70} & \textit{.76} & \textit{.73} & \textit{.70} & \textit{.74} & \textit{.71} & \textit{\textbf{.79}} & \textit{.78} & \textit{.77} & \textit{.75} & \textit{\textbf{.82}} & \textbf{\textit{.78}} & \textit{.69} & \textit{.78} & \textit{.72} \\
\multicolumn{2}{r}{\multirow{2}{*}{\textit{Difference with the within-platform setting}}} & \textit{Micro-avg.} & \textit{-.05} & \textit{-.05} & \textit{-.05} & \textit{-.05} & \textit{-.05} & \textit{-.05} & \textit{-.02} & \textit{-.02} & \textit{-.02} & \multicolumn{3}{|c}{\multirow{2}{*}{\textit{--}}} & \multicolumn{3}{|c}{\multirow{2}{*}{\textit{--}}} \\
\multicolumn{2}{r}{} & \textit{Macro-avg.} & \textit{-.09} & \textit{--} & \textit{-.04} & \textit{-.09} & \textit{-.02} & \textit{-.06} & \textit{-.04} & \textit{--} & \textit{-.03} & \multicolumn{3}{|c}{} & \multicolumn{3}{|c}{} \\
\hline
\multicolumn{1}{l}{\multirow{5}{*}{Cross-platform}} & \multicolumn{1}{l}{\multirow{5}{*}{Stack Overflow}} & Negative & .44 & .26 & .33 & .50 & .69 & .58 & .16 & .03 & .05 & \textbf{.64} & \textbf{.72} & \textbf{.68} & .52 & .69 & .59 \\
\multicolumn{1}{l}{} & \multicolumn{1}{l}{} & Neutral & .83 & .79 & .81 & .92 & .73 & .81 & .75 & \textbf{.94} & .83 & .93 & .81 & \textbf{.87} & \textbf{.96} & .75 & .83 \\
\multicolumn{1}{l}{} & \multicolumn{1}{l}{} & Positive & .57 & .85 & .68 & .63 & .90 & .74 & .72 & .49 & .58 & \textbf{.69} & \textbf{.93} & \textbf{.79} & .63 & .90 & .74 \\
\multicolumn{1}{l}{} & \multicolumn{1}{l}{} & \textit{Micro-avg.} & \textit{.73} & \textit{.73} & \textit{.73} & \textit{.76} & \textit{.76} & \textit{.76} & \textit{.73} & \textit{.73} & \textit{.73} & \textit{\textbf{.82}} & \textit{\textbf{.82}} & \textit{\textbf{.82}} & \textit{.77} & \textit{.77} & \textit{.77} \\
\multicolumn{1}{l}{} & \multicolumn{1}{l}{} & \textit{Macro-avg.} & \textit{.62} & \textit{.63} & \textit{.61} & \textit{.68} & \textit{.78} & \textit{.71} & \textit{.54} & \textit{.49} & \textit{.49} & \textit{\textbf{.75}} & \textit{\textbf{.82}} & \textit{\textbf{.78}} & \textit{.69} & \textit{.78} & \textit{.72} \\
\multicolumn{2}{r}{\multirow{2}{*}{\textit{Differences with the within-platform setting}}} & \textit{Micro-avg.} & \textit{-.10} & \textit{-.10} & \textit{-.10} & \textit{-.07} & \textit{-.07} & \textit{-.07} & \textit{-.13} & \textit{-.13} & \textit{-.13} & \multicolumn{3}{|c}{\multirow{2}{*}{--}} & \multicolumn{3}{|c}{\multirow{2}{*}{--}} \\
\multicolumn{2}{r}{} & \textit{Macro-avg.} & \textit{-.17} & \textit{-.13} & \textit{-.16} & \textit{-.11} & \textit{+.02} & \textit{-.06} & \textit{-.29} & \textit{-.29} & \textit{-.31} & \multicolumn{3}{|c}{} & \multicolumn{3}{|c}{} \\ 
\hline
\end{tabular}%
}
\end{table*} 

\begin{table}[h!]
\caption{Agreement of SE-specific tools with manual labelling. The within-platform setting for each dataset is reported in gray.}
\label{tab:kappamanual}
\resizebox{\columnwidth}{!}{%
\begin{tabular}{c|l|l|rrr}
\multirow{3}{*}{\textbf{Train set}} & \multicolumn{1}{c}{\multirow{3}{*}{\textbf{Classifier}}} & \multicolumn{4}{c}{\textbf{Agreement metrics}} \\
 & \multicolumn{1}{c}{} & \multicolumn{1}{c}{\multirow{2}{*}{\textit{k}}} & \multicolumn{1}{c}{\textit{Perfect}} & \multicolumn{2}{c}{\textit{Disagreement}} \\
 & \multicolumn{1}{c}{} & \multicolumn{1}{c}{} & \multicolumn{1}{c}{\textit{Agreement}} & \multicolumn{1}{c}{\textit{Severe}} & \multicolumn{1}{c}{\textit{Mild}} \\ \hline
\multicolumn{6}{|c|}{\textit{Test set: GitHub}} \\ \hline
\multirow{2}{*}{none} & Senti-Strength-SE & .71 & 80\% & 4\% & 16\% \\
 & DEVA & .58 & 73\% & 9\% & 18\% \\
\hline
\multirow{2}{*}{\begin{tabular}[c]{@{}c@{}}Stack \\ Overflow\end{tabular}} & Senti4SD & .61 & 74\% & 5\% & 21\% \\
 & SentiCR & .53 & 68\% & 3\% & 29\% \\
\hline
\multirow{2}{*}{Jira} & Senti4SD & .52 & 68\% & 2\% & 30\% \\
 & SentiCR & .35 & 58\% & 1\% & 41\% \\
\hline
\multirow{2}{*}{GitHub} & \cellcolor[HTML]{EFEFEF}Senti4SD & \cellcolor[HTML]{EFEFEF}.88 & \cellcolor[HTML]{EFEFEF}91\% & \cellcolor[HTML]{EFEFEF}1\% & \cellcolor[HTML]{EFEFEF}8\% \\
 & \cellcolor[HTML]{EFEFEF}SentiCR & \cellcolor[HTML]{EFEFEF}.74 & \cellcolor[HTML]{EFEFEF}83\% & \cellcolor[HTML]{EFEFEF}2\% & \cellcolor[HTML]{EFEFEF}15\% \\ \hline
\multicolumn{6}{|c|}{\textit{Test set: Stack Overflow}} \\ \hline
\hline
\multirow{2}{*}{none} & Senti-Strength-SE & .74 & 80\% & 2\% & 18\% \\
 & DEVA & .71 & 79\% & 4\% & 17\% \\
\hline
\multirow{2}{*}{GitHub} & Senti4SD & .69 & 77\% & 3\% & 20\% \\
 & SentiCR & .58 & 72\% & 5\% & 23\% \\
\hline
\multirow{2}{*}{Jira} & Senti4SD & .55 & 68\% & 1\% & 31\% \\
 & SentiCR & .31 & 54\% & 1\% & 45\% \\
\hline
\multirow{2}{*}{\begin{tabular}[c]{@{}c@{}}Stack \\ Overflow\end{tabular}} & \cellcolor[HTML]{EFEFEF}Senti4SD & \cellcolor[HTML]{EFEFEF}.83 & \cellcolor[HTML]{EFEFEF}87\% & \cellcolor[HTML]{EFEFEF}1\% & \cellcolor[HTML]{EFEFEF}12\% \\
 & \cellcolor[HTML]{EFEFEF}SentiCR & \cellcolor[HTML]{EFEFEF}.76 & \cellcolor[HTML]{EFEFEF}82\% & \cellcolor[HTML]{EFEFEF}3\% & \cellcolor[HTML]{EFEFEF}15\% \\ \hline
\multicolumn{6}{|c|}{\textit{Test set: Jira}} \\ \hline
\hline
\multirow{2}{*}{none} & Senti-Strength-SE & .69 & 82\% & 1\% & 17\% \\
 & DEVA & .6 & 77\% & 2\% & 21\% \\
\hline
\multirow{2}{*}{GitHub} & Senti4SD & .61 & 78\% & 1\% & 21\% \\
 & SentiCR & .68 & 84\% & 1\% & 16\% \\
\hline
\multirow{2}{*}{\begin{tabular}[c]{@{}c@{}}Stack \\ Overflow\end{tabular}} & Senti4SD & .47 & 73\% & 2\% & 25\% \\
 & SentiCR & .33 & 73\% & 2\% & 25\% \\
\hline
\multirow{2}{*}{Jira} & \cellcolor[HTML]{EFEFEF}Senti4SD & \cellcolor[HTML]{EFEFEF}.68 & \cellcolor[HTML]{EFEFEF}83\% & \cellcolor[HTML]{EFEFEF}-- & \cellcolor[HTML]{EFEFEF}17\% \\
 & \cellcolor[HTML]{EFEFEF}SentiCR & \cellcolor[HTML]{EFEFEF}.72 & \cellcolor[HTML]{EFEFEF}86\% & \cellcolor[HTML]{EFEFEF}-- & \cellcolor[HTML]{EFEFEF}14\% \\
 \hline
\end{tabular}%
}
\vspace{-3mm}
\end{table}

In Table 2, we report the performance in the cross-platform setting of the four tools, both by polarity class and overall. In bold we highlight the best values for each metric. For the sake of comparison against the within-platform setting, we also report the performance obtained by replicating the our previous study~\cite{Novielli18} (reported in grey). For each dataset, we highlighted in Italic the differences with respect to the within-platform setting. Furthermore, we report the tool agreement with the manual labeling (see Table~\ref{tab:kappamanual}) in terms of both weighted Cohen $\kappa$ and the percentage of cases in which each tool issues the correct prediction (perfect agreement with the gold label) as well as the percentage of wrong predictions (severe/mild disagreements). 

In the cross-platform setting, we observe a drop in the performance of the supervised tools SentiCR and Senti4SD on all datasets, compared to the within-platform setting. Conversely to what is observed in the within-platform setting, the two lexicon-based tools outperform the supervised approaches when these are retrained in a cross-platform condition. Exceptions are the cross-platform setting with training performed on Stack Overflow and test on GitHub, where Senti4SD achieve the best performance (macro F1 =.82), and the setting with training performed on GitHub and test on Stack Overflow, where Senti4SD and SentiStrength-SE both achieve the best performance (macro F1 = .80). The highest drop in performance is observed for SentiCR when Jira is used for training and GitHub (macro F1 = .52, representing a drop of 30\% with respect to the within-platform setting) and Stack Overflow for testing (macro F1 = .46, indicating a drop of 36\%). As a further confirmation of the results in Table 2, we observe a substantial agreement with the manual annotation for SentiStrength-SE (see Table~\ref{tab:kappamanual}). Conversely, the $\kappa$ values in the cross-platform setting indicate a moderate to substantial agreement for Senti4SD and DEVA, and a fair to moderate agreement for SentiCR.  

As for Senti4SD, a slight increase in performance is reported when BoW is excluded from the feature set in  most settings (see Table 2). For the GitHub test set, the macro F1 of Senti4SD raises from .80 (with BoW) to .82 (without BoW) when training on Stack Overflow, and from .66 (with BoW) to .69 (without BoW) when training on Jira. We observe similar results for Stack Overflow when GitHub is used to train and for the Jira test set with training on Stack Overflow. Consistently, we observe the highest drop in macro-average from the within- to the cross-platform setting for SentiCR (-30\%, -36\%, and -31\% decrease in macro F1 for the GitHub, Stack Overflow, and Jira test sets, respectively), which exploits a fixed feature set composed on uni- and bi-grams. This provides evidence of the lower ability to generalize of BoW features in cross-dataset settings, thus confirming the concerns of the NLP community about the risk of overfitting of model relying on n-gram features~\cite{Jurafsky_book}. 
Looking at the performance of each polarity class, we observe that the drop in performance is mainly due to a drop in precision for the neutral class and recall for the negative and positive classes. This evidence suggests that positive and negative lexicon might be platform-dependent and, therefore, we lose recall for the non-neutral classes in cross-platform settings. This also reflects in the mild disagreement (i.e., the confounding between the positive and neutral, or between the negative and neutral classes) being the main cause of disagreement. Conversely, severe disagreement occurs at most in the 9\% of cases, for DEVA on GitHub (see Table~\ref{tab:kappamanual}).

\paragraph{\textbf{RQ2} - To what extent do results from different SE-specific sentiment analysis tools agree with each other when used as ‘off-the-shelf’ tools in a cross-platform setting?} - In Table~\ref{tab:paired_kappa}, we report the paired comparisons, using the same measures of agreement between each pair of tools. SentiStrength-SE and DEVA also show a substantial to almost perfect agreement with each other, ranging from $\kappa =.65$ for GitHub to $\kappa = .79$ for Stack Overflow, and $\kappa = .81$ for Jira. This is somewhat expected, considering that they share the same lexical resources and approach for polarity classification~\cite{Islam17,Islam18}. The lowest agreement scores are observed for the lexicon-based tools and SentiCR, which is purely based on BoW. Senti4SD is in the middle of this scale, showing a moderate to substantial agreement with lexicon-based tools, probably because it relies on both lexicon-based features and BoW.

\begin{table}[h!]
\caption{Agreement of SE-specific tools with each other in cross-platform settings. The within-platform setting for each dataset is reported in gray.}
\label{tab:paired_kappa}
\resizebox{\columnwidth}{!}{%
\begin{tabular}{c|l|l|rrr}
\multicolumn{1}{l|}{\multirow{3}{*}{\textbf{Train set}}} & \multirow{3}{*}{\textbf{Classifier}} & \multicolumn{4}{c}{\textbf{Agreement metrics}} \\ \cline{3-6} 
\multicolumn{1}{l|}{} &  & \multirow{2}{*}{\textit{k}} & \multicolumn{1}{l}{\textit{Perfect}} & \multicolumn{2}{l}{\textit{Disagreement}} \\
\multicolumn{1}{l|}{} &  &  & \multicolumn{1}{l}{\textit{Agreement}} & \multicolumn{1}{l}{\textit{Severe}} & \multicolumn{1}{l}{\textit{Mild}} \\ \hline
\multicolumn{6}{|c|}{\textit{Test set: GitHub}} \\ \hline
\multicolumn{1}{l|}{--} & SentiStrength-SE vs DEVA & 0.65 & 78\% & 7\% & 15\% \\ \hline
\multirow{5}{*}{\begin{tabular}[c]{@{}c@{}}Stack \\ Overflow\end{tabular}} & Senti4SD vs. SentiCR & 0.48 & 68\% & 3\% & 29\% \\
 & Senti4SD vs. SentiStrength-SE & 0.58 & 73\% & 5\% & 22\% \\
 & Senti4SD vs. DEVA & 0.46 & 65\% & 8\% & 27\% \\
 & SentiCR vs. SentiStrength-SE & 0.51 & 68\% & 3\% & 29\% \\
 & SentiCR vs. DEVA & 0.47 & 64\% & 5\% & 31\% \\ \hline
\multirow{5}{*}{Jira} & Senti4SD vs. SentiCR & 0.49 & 78\% & 0\% & 22\% \\
 & Senti4SD vs. SentiStrength-SE & 0.51 & 69\% & 2\% & 29\% \\
 & Senti4SD vs. DEVA & 0.38 & 58\% & 4\% & 38\% \\
 & SentiCR vs. SentiStrength-SE & 0.39 & 63\% & 1\% & 36\% \\
 & SentiCR vs. DEVA & 0.30 & 53\% & 2\% & 45\% \\ \hline
\multirow{5}{*}{GitHub} & \cellcolor[HTML]{EFEFEF}Senti4SD vs. SentiCR & \cellcolor[HTML]{EFEFEF}0.75 & \cellcolor[HTML]{EFEFEF}83\% & \cellcolor[HTML]{EFEFEF}2\% & \cellcolor[HTML]{EFEFEF}15\% \\
 & \cellcolor[HTML]{EFEFEF}Senti4SD vs. SentiStrength-SE & \cellcolor[HTML]{EFEFEF}0.71 & \cellcolor[HTML]{EFEFEF}81\% & \cellcolor[HTML]{EFEFEF}4\% & \cellcolor[HTML]{EFEFEF}15\% \\
 & \cellcolor[HTML]{EFEFEF}Senti4SD vs. DEVA & \cellcolor[HTML]{EFEFEF}0.59 & \cellcolor[HTML]{EFEFEF}73\% & \cellcolor[HTML]{EFEFEF}8\% & \cellcolor[HTML]{EFEFEF}19\% \\
 & \cellcolor[HTML]{EFEFEF}SentiCR vs. SentiStrength-SE & \cellcolor[HTML]{EFEFEF}0.65 & \cellcolor[HTML]{EFEFEF}77\% & \cellcolor[HTML]{EFEFEF}4\% & \cellcolor[HTML]{EFEFEF}19\% \\
 & \cellcolor[HTML]{EFEFEF}SentiCR vs. DEVA & \cellcolor[HTML]{EFEFEF}0.56 & \cellcolor[HTML]{EFEFEF}72\% & \cellcolor[HTML]{EFEFEF}7\% & \cellcolor[HTML]{EFEFEF}21\% \\ \hline
\multicolumn{6}{|c|}{\textit{Test set:  Stack Overflow}} \\ \hline
\multicolumn{1}{l|}{--} & SentiStrength-SE vs DEVA & 0.79 & 85\% & 4\% & 11\% \\ \hline
\multirow{5}{*}{GitHub} & Senti4SD vs. SentiCR & 0.59 & 73\% & 5\% & 22\% \\
 & Senti4SD vs. SentiStrength-SE & 0.69 & 76\% & 2\% & 22\% \\
 & Senti4SD vs. DEVA & 0.64 & 74\% & 5\% & 21\% \\
 & SentiCR vs. SentiStrength-SE & 0.57 & 70\% & 4\% & 26\% \\
 & SentiCR vs. DEVA & 0.55 & 69\% & 7\% & 24\% \\ \hline
\multirow{5}{*}{Jira} & Senti4SD vs. SentiCR & 0.44 & 74\% & 1\% & 25\% \\
 & Senti4SD vs. SentiStrength-SE & 0.53 & 65\% & 1\% & 34\% \\
 & Senti4SD vs. DEVA & 0.49 & 62\% & 2\% & 36\% \\
 & SentiCR vs. SentiStrength-SE & 0.33 & 55\% & 1\% & 44\% \\
 & SentiCR vs. DEVA & 0.29 & 49\% & 1\% & 50\% \\ \hline
\multirow{5}{*}{Stack Overflow} & \cellcolor[HTML]{EFEFEF}Senti4SD vs. SentiCR & \cellcolor[HTML]{EFEFEF}0.75 & \cellcolor[HTML]{EFEFEF}82\% & \cellcolor[HTML]{EFEFEF}4\% & \cellcolor[HTML]{EFEFEF}14\% \\
 & \cellcolor[HTML]{EFEFEF}Senti4SD vs. SentiStrength-SE & \cellcolor[HTML]{EFEFEF}0.79 & \cellcolor[HTML]{EFEFEF}83\% & \cellcolor[HTML]{EFEFEF}2\% & \cellcolor[HTML]{EFEFEF}15\% \\
 & \cellcolor[HTML]{EFEFEF}Senti4SD vs. DEVA & \cellcolor[HTML]{EFEFEF}0.73 & \cellcolor[HTML]{EFEFEF}80\% & \cellcolor[HTML]{EFEFEF}5\% & \cellcolor[HTML]{EFEFEF}15\% \\
 & \cellcolor[HTML]{EFEFEF}SentiCR vs. SentiStrength-SE & \cellcolor[HTML]{EFEFEF}0.72 & \cellcolor[HTML]{EFEFEF}80\% &\cellcolor[HTML]{EFEFEF}4\% & \cellcolor[HTML]{EFEFEF}16\% \\
 & \cellcolor[HTML]{EFEFEF}SentiCR vs. DEVA & \cellcolor[HTML]{EFEFEF}0.68 & \cellcolor[HTML]{EFEFEF}79\% & \cellcolor[HTML]{EFEFEF}7\% &\cellcolor[HTML]{EFEFEF}14\% \\ \hline
\multicolumn{6}{|c|}{\textit{Test set: Jira}} \\ \hline
\multicolumn{1}{l|}{--} & SentiStrength-SE vs DEVA & 0.81 & 90\% & 3\% & 7\% \\ \hline
\multirow{5}{*}{GitHub} & Senti4SD vs. SentiCR & 0.71 & 84\% & 1\% & 15\% \\
 & Senti4SD vs. SentiStrength-SE & 0.71 & 83\% & 2\% & 15\% \\
 & Senti4SD vs. DEVA & 0.63 & 79\% & 3\% & 18\% \\
 & SentiCR vs. SentiStrength-SE & 0.76 & 87\% & 2\% & 11\% \\
 & SentiCR vs. DEVA & 0.69 & 83\% & 3\% & 14\% \\ \hline
\multirow{5}{*}{\begin{tabular}[c]{@{}c@{}}Stack \\ Overflow\end{tabular}} & Senti4SD vs. SentiCR & 0.38 & 74\% & 1\% & 24\% \\
 & Senti4SD vs. SentiStrength-SE & 0.61 & 79\% & 3\% & 18\% \\
 & Senti4SD vs. DEVA & 0.54 & 75\% & 4\% & 21\% \\
 & SentiCR vs. SentiStrength-SE & 0.33 & 69\% & 2\% & 29\% \\
 & SentiCR vs. DEVA & 0.29 & 65\% & 3\% & 32\% \\ \hline
\multirow{5}{*}{\textit{Jira}} & \cellcolor[HTML]{EFEFEF}Senti4SD vs. SentiCR & \cellcolor[HTML]{EFEFEF}0.77 & \cellcolor[HTML]{EFEFEF}89\% & \cellcolor[HTML]{EFEFEF}0\% & \cellcolor[HTML]{EFEFEF}11\% \\
 & \cellcolor[HTML]{EFEFEF}Senti4SD vs. SentiStrength-SE & \cellcolor[HTML]{EFEFEF}0.70 & \cellcolor[HTML]{EFEFEF}84\% & \cellcolor[HTML]{EFEFEF}1\% & \cellcolor[HTML]{EFEFEF}15\% \\
 & \cellcolor[HTML]{EFEFEF}Senti4SD vs. DEVA & \cellcolor[HTML]{EFEFEF}0.63 & \cellcolor[HTML]{EFEFEF}79\% & \cellcolor[HTML]{EFEFEF}2\% & \cellcolor[HTML]{EFEFEF}19\% \\
 & \cellcolor[HTML]{EFEFEF} SentiCR vs. SentiStrength-SE & \cellcolor[HTML]{EFEFEF}0.76 & \cellcolor[HTML]{EFEFEF}87\% & \cellcolor[HTML]{EFEFEF}1\% & \cellcolor[HTML]{EFEFEF}12\% \\
 & \cellcolor[HTML]{EFEFEF}SentiCR vs. DEVA & \cellcolor[HTML]{EFEFEF}0.69 & \cellcolor[HTML]{EFEFEF}82\% & \cellcolor[HTML]{EFEFEF}1\% & \cellcolor[HTML]{EFEFEF}17\%\\
\hline
\end{tabular}%
}
\vspace{-3mm}
\end{table}

\subsection{Error Analysis}
\label{sec:error-analysis}
\begin{table}[ht]
\caption{Distribution of error categories}
\label{tab:errors}
\begin{tabular}{lr}
\textbf{Error category} & \textbf{\#cases (\%)} \\
\hline
General error & 214 (68\%) \\
Subjectivity in annotation & 35 (11\%) \\
Polar facts & 25 (8\%) \\
Politeness & 19 (6\%) \\
Implicit sentiment polarity & 16 (5\%) \\
Figurative language & 6 (2\%) \\
Pragmatics & 6 (2\%) \\
\hline
Overall & 320
\end{tabular}%
\vspace{-3mm}
\end{table}

We complement the quantitative analysis with a content analysis  aimed at assessing the main causes of misclassification. We randomly sampled a subset of 320 texts (statistically significant sample size at 95\% confidence level) from the documents for which both supervised classifiers yield a wrong prediction. Two of the authors independently labeled half of the cases and assigned a label choosing among the error categories identified in our previous benchmark study~\cite{Novielli18} (see Table~\ref{tab:errors}). Then, they jointly discussed all cases to confirm the error labels. The goal of this analysis is to assess if the open challenges in sentiment analysis of developers’ communication traces in a cross-platform setting are the same highlighted in a within-platform condition.

We found that the main cause of misclassification are \textit{general errors}, occurring 68\% of times. Such errors are caused by the inability of the tools to correctly deal with some textual cues. In most cases, this is due to lexical cues  that are not recognized as either positive or negative because they do not occur frequently enough in the train set in order to hold sufficient predictive power. A special case is emoticons, which may have platform-dependent representation (e.g., ":smiley:" vs. ":-)").  
General errors also occur due to wrong preprocessing (e.g., emoticons erroneously treated as non-unique tokens and rather split into its constituent characters), wrong spelling of words, or wrong negation handling. 

The second cause for misclassification is the \textit{subjectivity in sentiment annotation} (11\%). Sentiment labeling is an inherently subjective task: even in the presence of clear annotation guidelines, the label assigned to a given text might be influenced by the personality traits of the human annotator~\cite{Scherer2004}. In line with previous results~\cite{Novielli18}, we observe that in some cases, the raters are conservative and provide a neutral label for mild expressions of emotions or opinions. 

Furthermore, the specific research goal and applications of sentiment analysis might be another driver for labeling decisions. It is the case of \textit{polar facts}, which are inherently desirable or undesirable facts, such as code patch acceptance (e.g., "fixed") or bug reports (e.g., "seems to be failing for a different reason now"), expressed with a neutral sentiment. Polar facts are the third cause of misclassification in the cross-platform setting (8\%), as they might be inconsistently labeled across datasets, in line with the specific goals of the authors. For example, polar facts are often labeled as non-neutral in Jira. As an example, sentences such as "This seems to be failing for different reasons" or "This might be a bug indeed" are labeled as negative even if a neutral style is used (absence of emotions), probably due to the original intention of the authors of the Jira dataset to analyze the role of sentiment in issue tracking and its correlation with issue fixing time~\cite{Murgia2014}. Polar facts are reported as the main cause of error in the within-platform setting~\cite{Novielli18}.

The misclassification of sentences conveying \textit{politeness} is a cause of error in 6\% of cases, due to politeness expression such as "Thanks!" or "Sorry for" being inconsistently labeled across-dataset. As an example, in the Stack Overflow and GitHub datasets, politeness is considered neutral unless a clear expression of emotion is present in the text. This choice is in line with the evidence provided by computational linguists that emotion lexicon can be used for politeness expressions. This is typical of the so-called \textit{behabitives} speech acts~\cite{Austin1962}, in which no real feelings are expressed, but still emotional words are employed to convey other communicative intentions (e.g., "I am afraid this does not work"). As for Jira, thanking expression receive a positive label when they are related to code change approval (e.g., "thanks for the patch" is positive) indicating that positive polar facts receive a positive label (the patch is satisfying), while expression of gratitude (as in "Thanks!") are usually interpreted as neutral. Again, this is in line with the intention of Murgia et al. to study how sentiment correlates with issue-fixing time~\cite{Murgia2014}.

In 5\% of cases, the sentiment is conveyed through indirect lexicon (\textit{Implicit sentiment polarity}). As such, these comments are erroneously classified as neutral due to the absence of explicit lexical cues of sentiment. Finally, a few cases (2\%) are misclassified due to the inability of the classifiers to deal with \textit{figurative language}, as in the presence of humor or irony. The remaining 2\% of cases are misclassified because the classifiers are not designed to take into account \textit{pragmatics}. It is the case of questions or sentences reporting third persons' opinions or emotions, which are correctly labeled as neutral by humans but   misclassified by the tool as positive or negative due to the presence of emotion words.

\subsection{Learning curves for supervised classifiers}
\label{sec:learning-curves}
\begin{figure}
	\centering
	\begin{subfigure}{\columnwidth} 
		\includegraphics[width=\columnwidth]{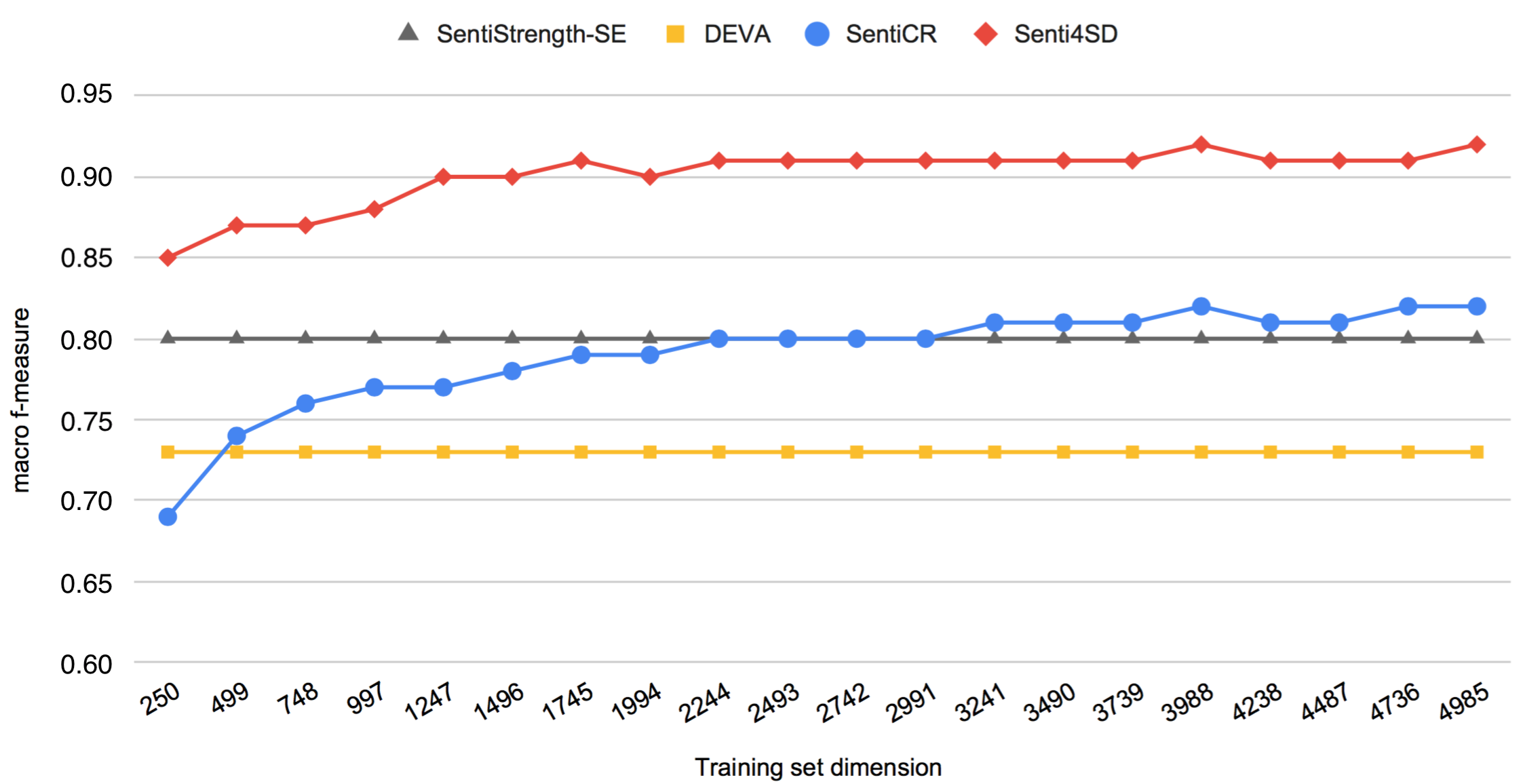}
		\caption{GitHub} 
		\vspace{1em} 
	\end{subfigure}
	\begin{subfigure}{\columnwidth} 
		\includegraphics[width=\columnwidth]{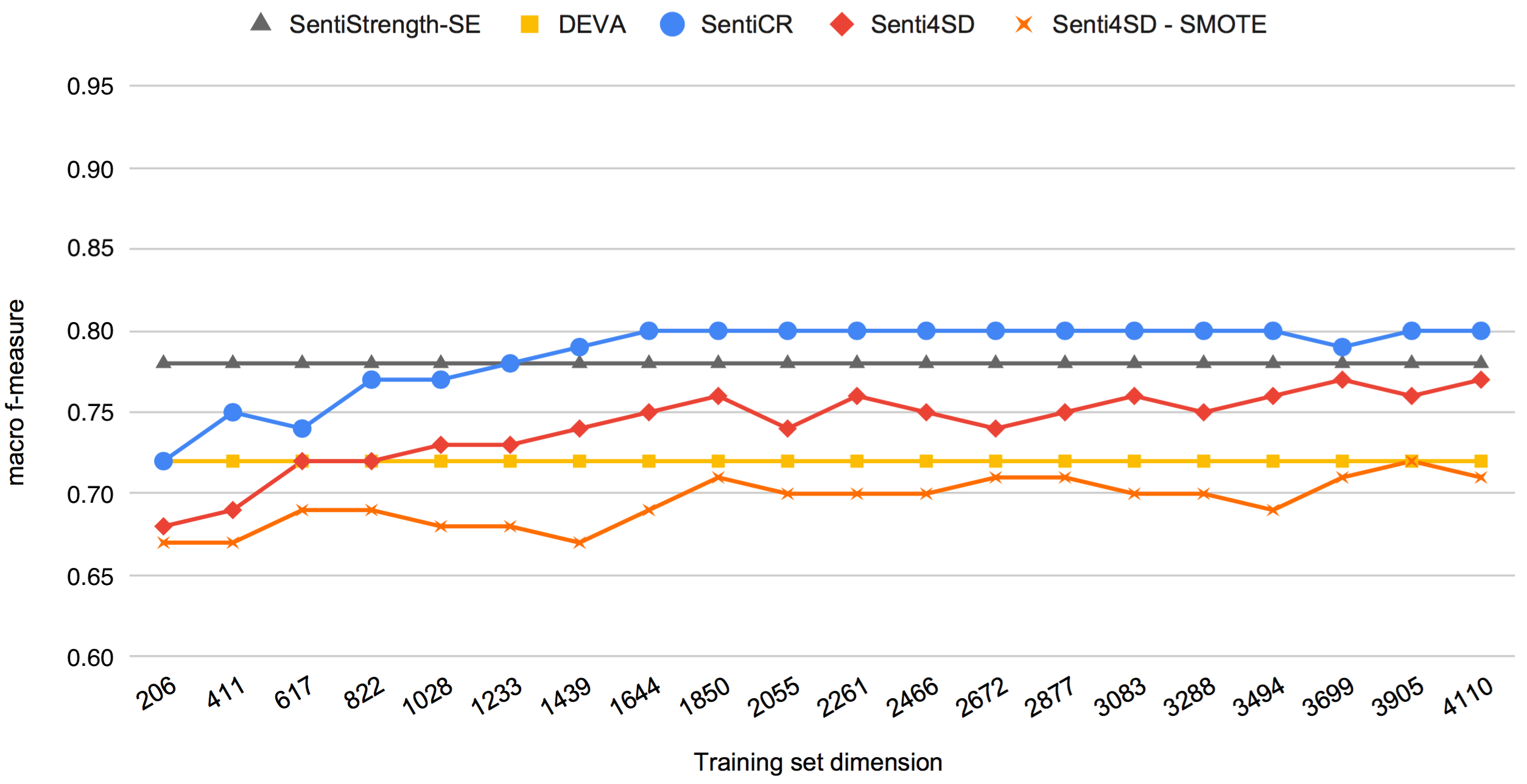}
		\caption{Jira} 
		\vspace{1em} 
	\end{subfigure}
	\begin{subfigure}{\columnwidth} 
		\includegraphics[width=\columnwidth]{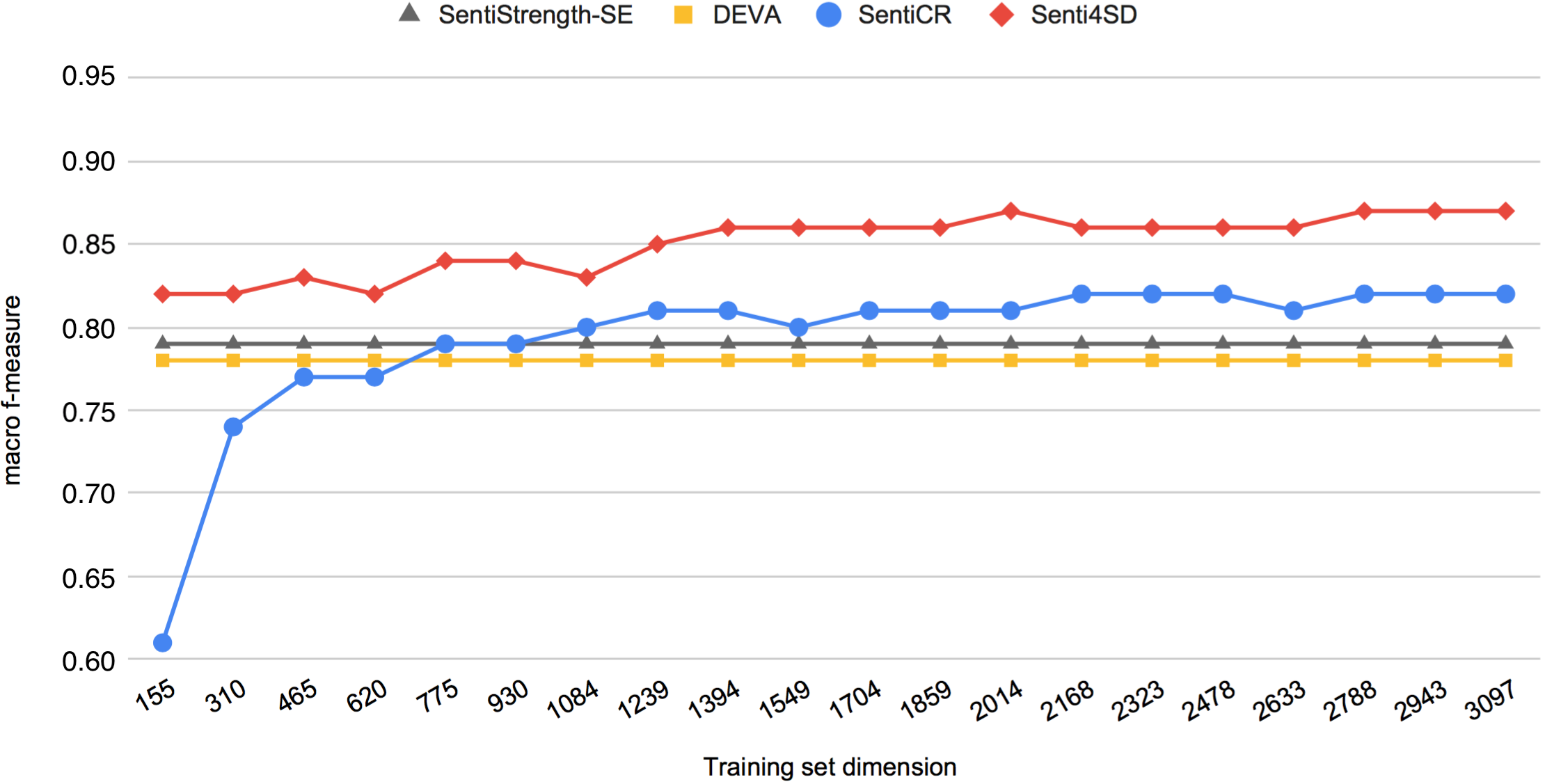}
		\caption{Stack Overflow} 
	\end{subfigure}
	\setlength{\belowcaptionskip}{-5mm}
	\caption{Learning curves for the supervised tools evaluated in a within-platform setting using the GitHub (a), Jira (b), and Stack Overflow (c) datasets.} 
\label{fig:learning_curves}	
\end{figure}

\paragraph{\textbf{RQ3} - To what extent is the performance of SE-specific sentiment analysis tools affected by the size of the training set?} - We want to assess how many documents we need to reliably retrain a supervised classifier for sentiment analysis in the software engineering domain. Accordingly, we analyze the learning curves of Senti4SD and SentiCR in a within-platform setting (see Figure~\ref{fig:learning_curves}). The performance of the lexicon-based tools, which cannot be customized, is reported for reference. We obtain the learning curves by plotting the performance on the test set of models created with training subsets of incremental size. We start by randomly sampling a subset of 5\% of the original train set, using stratified sampling to preserve the polarity label distribution. At each iteration, we increase the training set at a rate of 5\% and assess the model performance on the same 30\% held-out test set used to address RQ1 and RQ2.

We observe that for GitHub and Stack Overflow, retraining Senti4SD is always convenient, even with a minimal set of documents, compared to the performance of lexicon-based tools. 
The nearly-optimal performance is obtained, for both datasets, with a train set of about 1,200 documents. 
A different situation is observed for Jira, which is largely unbalanced in favor of the neutral class (67\% of the dataset). In this case, retraining is beneficial only if a larger set of documents is available (about 1,600 texts). For example, for SentiCR more than 1,200 documents are required to outperform SentiStrength-SE (see Figure~\ref{fig:learning_curves}.b). However, the improvement is negligible if compared to the one observed for GitHub (see Figure~\ref{fig:learning_curves}.a) and Stack Overflow (see Figure~\ref{fig:learning_curves}.c).
A possible explanation for these results is that SentiStrength-SE was originally optimized using a subset of the Jira gold standard as a reference~\cite{Islam17}, which may arguably explain its very good performance on it.
Another possible explanation for this difference in the performance could reside in the Jira dataset being unbalanced, thus making retraining not as effective as for GitHub and Stack Overflow, which are well-balanced datasets. As such, we included an additional setting for Jira where we performed class-balancing using SMOTE also for Senti4SD (SMOTE is the default preprocessing for SentiCR). This evidence suggests that even if resampling is performed before retraining, SentiStrength-SE still outperforms the other tools.
As a further possible explanation, we hypothesize that the quality of the gold standard, measured in terms of inter-rater agreement, is also a major fact influencing the quality and reliability of the learned classification model. In fact, for both GitHub and Stack Overflow, $\kappa$ values indicate a substantial to almost perfect agreement, while a lower agreement is observed for Jira (see Section~\ref{sec:datasets}).

\section{Discussion}
\label{sec:discussion}
In the following, we derive empirically-driven guidelines for reliable sentiment analysis in SE, based on the findings of the current study.

\textbf{Perform SE-specific tuning for enhanced accuracy}. Domain adaptation is a well-known problem in machine learning~\cite{David2010}, in general, and in sentiment analysis, in particular~\cite{Plank2018}. 
Our previous benchmarking study performed in a within-platform setting on the Stack Overflow and Jira datasets demonstrated that SE-specific tuning is beneficial for ensuring reliable sentiment analysis on technical texts~\cite{Novielli18}. 
We confirm these findings also on the GitHub dataset that we developed for the purpose of enriching the benchmark in the current study. In particular, we report comparable performance for the lexicon-based tools SentiStrength-SE and DEVA, thus providing further evidence that reliable sentiment analysis in software engineering is a feasible task.   

\textbf{Perform platform-specific tuning}. The results of our benchmark study demonstrate how retraining across platforms does not work well for supervised tools, thus suggesting that the definition of `domain’ might be even narrowed-down at the level of the specific platform. In fact, despite our benchmark included only SE-specific datasets, we observe a drop in performance when supervised models are trained and tested on data gathered from different collaborative development environments. This suggests that semantics shifts also occur due to platform-specific jargon and communication style. In line with this evidence, we report better performance in the absence of BoW-based features (i.e., for Senti4SD no BoW, see Table 2) indicating the lower ability of n-grams to generalize, i.e., they might cause overfitting to the platform-specific lexicon, thus negatively affecting the performance of supervised tools. This is further confirmed by the results of our error analysis (see Section~\ref{sec:error-analysis}).  As such, whenever a gold standard is available, we recommend platform-specific retraining to enable correct modeling of the interaction style and lexicon of the specific platform. 

\textbf{Build a robust gold standard}. In building a gold standard, one open issue is the correct amount of data required for retraining a reliable supervised classifier. To address this question, we performed a within-platform study and built the learning curves obtained with training sets of incremental size. The results, depicted in Figure~\ref{fig:learning_curves}, show that learning from unbalanced, low-agreement data might produce unsatisfying results even in a within-platform setting. This claim is in line with previous findings suggesting that the quality~\cite{Agrawal18,Tantithamthavorn15} and internal consistency~\cite{Novielli18} of gold standards are crucial properties for successful training of classifiers. 

\textbf{Select the appropriate tool in line with the research goals}. In the absence of a platform-specific gold standard for retraining, unsupervised tools or `off-the-shelf’ use of supervised classifiers are the only possible options. In both cases, we recommend using a tool only if a preliminary sanity check produces satisfying results on the target platform. Specifically, we recommend to collect and manually annotate sample data from the target platform in order to verify the alignment between the classification output and the manually-provided labels. Indeed, one of the most dangerous assumptions when reusing sentiment analysis tools and datasets is assuming agreement with the goals and sentiment conceptualization as originally thought by their authors.  
Our error analysis shows that even when sharing the theoretical model of emotion (e.g., the Shaver model used for the three datasets), the human raters may provide polarity labels based on their subjective perception or the specific research goals. It is the case of politeness, which is labeled inconsistently across datasets (see Section~\ref{sec:error-analysis}), thus inducing misclassification in the cross-platform settings.

\section{Threats to Validity}
\label{sec:threats}
We are aware that the methodology adopted could produce different results if applied to different datasets and, therefore, that the choice of datasets to include in the benchmark might represent a threat to conclusion validity. As such, we included all the model-driven gold standards for sentiment annotation in software engineering that are available at the time of writing, composed of posts (questions, answers, and comments) from Stack Overflow and comments from Jira. To further mitigate this threat, we built a third gold standard dataset including comments from GitHub.

All datasets in our benchmark are built by collecting documents from platforms that are popular and widely adopted among software developers. As such, we included three major collaborative software development platforms. Each platform supports different collaborative tasks, from technical question-answering (Stack Overflow) to issue tracking (Jira), to collaborative software development with version control (GitHub). Given the dataset size and the variety of tasks considered, we are reasonably confident that the datasets included in this study are representative of the developers’ communication, thus reducing threats to external validity.

A threat to construct validity is due to sentiment analysis being inherently affected by the subjectivity of the studied phenomenon, i.e., emotions and opinions as conveyed in text~\cite{Scherer2004}. In our previous research~\cite{Novielli18}, we showed how model-driven annotation is crucial to obtain a high-quality, reliable gold standard for training emotion polarity classifiers. Inconsistency in the annotation guidelines might be a cause of a drop in performance \textit{per se}. As such, we addressed this threat by including in our benchmark only model-driven datasets. Furthermore, the GitHub dataset, which we built from scratch, is annotated following the same guidelines and adopting the same theoretical model of emotions leveraged for creating the Stack Overflow and Jira gold standards. This choice reduces the risk of confounding factors due to different annotation schema, thus enabling us to correctly assess the impact of the cross-platform train-test condition. 

Finally, threats to internal validity concern internal factors such as the configuration of the parameters for the machine learning algorithms implemented by Senti4SD and SentiCR. To mitigate this threat, we replicated the experimental conditions under which the tools were originally validated~\cite{CalefatoEMSE18}, ~\cite{Ahmed17}, using the available training toolkits. Furthermore, we ran again the within-platform setting to enable a fair comparison with the results reported in our previous research~\cite{Novielli18}.

\section{Conclusions}
\label{sec:conclusions}
In this paper, we assessed the performance of four available SE-specific sentiment analysis tools in a cross-platform setting. We found that the retraining of SE-specific sentiment analysis tools is not a viable solution when the training and test sets come from different data sources. Conversely, better performance is observed for lexicon-based approaches, which we recommend whenever retraining is not possible due to the unavailability of a gold standard. 
However, further evidence shows that supervised tools achieve better performance than lexicon-based ones when retrained with a minimal training set of about 1,000 documents, as long as the training set is balanced and substantial inter-rater agreement is observed. 
Based on our empirical findings, we derived guidelines for reliable sentiment analysis in software engineering. Finally, we built a dataset of over 7,000 manually annotated GitHub comments, which we release to support future studies in the field. 

In future work, we plan to further enhance the understanding of classification performance drop under domain- and platform-shift, by including the assessment of predictive power of features across additional datasets. Also, we plan to assess the cross-platform performance of approaches based on deep learning, which are not included in this study. 

\section{Acknowledgments}
We thank Giovanna Saracino for contributing to the early stage of this study. 

\bibliographystyle{ACM-Reference-Format}
\bibliography{bibliography}


\end{document}